\documentclass[12pt,fleqn]{book} 

\usepackage[top=3cm,bottom=3cm,left=3cm,right=3cm,headsep=10pt,a4paper]{geometry} 

\usepackage{graphicx} 
\graphicspath{{Pictures/}} 

\usepackage{lipsum} 

\usepackage{tikz} 

\usepackage[english]{babel} 

\usepackage{enumitem} 
\setlist{nolistsep} 

\usepackage{booktabs} 

\usepackage{xcolor} 
\definecolor{ocre}{RGB}{243,102,25} 

\usepackage{avant} 
\usepackage{mathptmx} 

\usepackage{microtype} 
\usepackage[utf8]{inputenc} 
\usepackage[T1]{fontenc} 

\usepackage{csquotes}
\usepackage[style=numeric,citestyle=numeric,sorting=nyt,sortcites=true,autopunct=true,autolang=hyphen,hyperref=true,abbreviate=false,backref=true,backend=bibtex,defernumbers=true]{biblatex}
\defbibheading{bibempty}{}

\usepackage{calc} 
\usepackage{makeidx} 
\makeindex 

\usepackage{titletoc} 

\contentsmargin{0cm} 

\titlecontents{part}[0cm]
{\addvspace{20pt}\centering\large\bfseries}
{}
{}
{}

\titlecontents{chapter}[1.25cm] 
{\addvspace{12pt}\large\sffamily\bfseries} 
{\color{ocre!60}\contentslabel[\Large\thecontentslabel]{1.25cm}\color{ocre}} 
{\color{ocre}}  
{\color{ocre!60}\normalsize\;\titlerule*[.5pc]{.}\;\thecontentspage} 

\titlecontents{section}[1.25cm] 
{\addvspace{3pt}\sffamily\bfseries} 
{\contentslabel[\thecontentslabel]{1.25cm}} 
{}
{\hfill\color{black}\thecontentspage} 
[]

\titlecontents{subsection}[1.25cm] 
{\addvspace{1pt}\sffamily\small} 
{\contentslabel[\thecontentslabel]{1.25cm}} 
{}
{\ \titlerule*[.5pc]{.}\;\thecontentspage} 
[]

\titlecontents{figure}[0em]
{\addvspace{-5pt}\sffamily}
{\thecontentslabel\hspace*{1em}}
{}
{\ \titlerule*[.5pc]{.}\;\thecontentspage}
[]

\titlecontents{table}[0em]
{\addvspace{-5pt}\sffamily}
{\thecontentslabel\hspace*{1em}}
{}
{\ \titlerule*[.5pc]{.}\;\thecontentspage}
[]

\titlecontents{lchapter}[0em] 
{\addvspace{15pt}\large\sffamily\bfseries} 
{\color{ocre}\contentslabel[\Large\thecontentslabel]{1.25cm}\color{ocre}} 
{}  
{\color{ocre}\normalsize\sffamily\bfseries\;\titlerule*[.5pc]{.}\;\thecontentspage} 

\titlecontents{lsection}[0em] 
{\sffamily\small} 
{\contentslabel[\thecontentslabel]{1.25cm}} 
{}
{}

\titlecontents{lsubsection}[.5em] 
{\normalfont\footnotesize\sffamily} 
{}
{}
{}

\usepackage{fancyhdr} 

\fancypagestyle{plain}{\fancyhead{}} 

\makeatletter
\renewcommand{\cleardoublepage}{
\clearpage\ifodd\c@page\else
\hbox{}
\vspace*{\fill}
\thispagestyle{empty}
\newpage
\fi}

\usepackage{amsmath,amsfonts,amssymb,amsthm} 

\newtheoremstyle{ocrenumbox}
{0pt}
{0pt}
{\normalfont}
{}
{\small\bf\sffamily\color{ocre}}
{\;}
{0.25em}
{\small\sffamily\color{ocre}\thmname{#1}\nobreakspace\thmnumber{\@ifnotempty{#1}{}\@upn{#2}}
\thmnote{\nobreakspace\the\thm@notefont\sffamily\bfseries\color{black}---\nobreakspace#3.}} 

\newtheoremstyle{blacknumex}
{5pt}
{5pt}
{\normalfont}
{} 
{\small\bf\sffamily}
{\;}
{0.25em}
{\small\sffamily{\tiny\ensuremath{\blacksquare}}\nobreakspace\thmname{#1}\nobreakspace\thmnumber{\@ifnotempty{#1}{}\@upn{#2}}
\thmnote{\nobreakspace\the\thm@notefont\sffamily\bfseries---\nobreakspace#3.}}

\newtheoremstyle{blacknumbox} 
{0pt}
{0pt}
{\normalfont}
{}
{\small\bf\sffamily}
{\;}
{0.25em}
{\small\sffamily\thmname{#1}\nobreakspace\thmnumber{\@ifnotempty{#1}{}\@upn{#2}}
\thmnote{\nobreakspace\the\thm@notefont\sffamily\bfseries---\nobreakspace#3.}}

\newtheoremstyle{ocrenum}
{5pt}
{5pt}
{\normalfont}
{}
{\small\bf\sffamily\color{ocre}}
{\;}
{0.25em}
{\small\sffamily\color{ocre}\thmname{#1}\nobreakspace\thmnumber{\@ifnotempty{#1}{}\@upn{#2}}
\thmnote{\nobreakspace\the\thm@notefont\sffamily\bfseries\color{black}---\nobreakspace#3.}} 
\makeatother

\newcounter{dummy} 
\numberwithin{dummy}{section}
\theoremstyle{ocrenumbox}
\newtheorem{theoremeT}[dummy]{Theorem}

\newtheorem{exerciseT}{Exercise}[chapter]
\theoremstyle{blacknumex}
\newtheorem{exampleT}{Example}[chapter]
\theoremstyle{blacknumbox}
\newtheorem{vocabulary}{Vocabulary}[chapter]
\newtheorem{definitionT}{Definition}[section]
\newtheorem{corollaryT}[dummy]{Corollary}
\theoremstyle{ocrenum}

\RequirePackage[framemethod=default]{mdframed} 

\newmdenv[skipabove=7pt,
skipbelow=7pt,
backgroundcolor=black!5,
linecolor=ocre,
innerleftmargin=5pt,
innerrightmargin=5pt,
innertopmargin=5pt,
leftmargin=0cm,
rightmargin=0cm,
innerbottommargin=5pt]{tBox}

\newmdenv[skipabove=7pt,
skipbelow=7pt,
rightline=false,
leftline=true,
topline=false,
bottomline=false,
backgroundcolor=ocre!10,
linecolor=ocre,
innerleftmargin=5pt,
innerrightmargin=5pt,
innertopmargin=5pt,
innerbottommargin=5pt,
leftmargin=0cm,
rightmargin=0cm,
linewidth=4pt]{eBox}	

\newmdenv[skipabove=7pt,
skipbelow=7pt,
rightline=false,
leftline=true,
topline=false,
bottomline=false,
linecolor=ocre,
innerleftmargin=5pt,
innerrightmargin=5pt,
innertopmargin=0pt,
leftmargin=0cm,
rightmargin=0cm,
linewidth=4pt,
innerbottommargin=0pt]{dBox}	

\newmdenv[skipabove=7pt,
skipbelow=7pt,
rightline=false,
leftline=true,
topline=false,
bottomline=false,
linecolor=gray,
backgroundcolor=black!5,
innerleftmargin=5pt,
innerrightmargin=5pt,
innertopmargin=5pt,
leftmargin=0cm,
rightmargin=0cm,
linewidth=4pt,
innerbottommargin=5pt]{cBox}

\newenvironment{exercise}{\begin{eBox}\begin{exerciseT}}{\hfill{\color{ocre}\tiny\ensuremath{\blacksquare}}\end{exerciseT}\end{eBox}}

\makeatletter
\renewcommand{\@seccntformat}[1]{\llap{\textcolor{ocre}{\csname the#1\endcsname}\hspace{1em}}}                    
\renewcommand{\section}{\@startsection{section}{1}{\z@}
{-4ex \@plus -1ex \@minus -.4ex}
{1ex \@plus.2ex }
{\normalfont\large\sffamily\bfseries}}
\renewcommand{\subsection}{\@startsection {subsection}{2}{\z@}
{-3ex \@plus -0.1ex \@minus -.4ex}
{0.5ex \@plus.2ex }
{\normalfont\sffamily\bfseries}}
\renewcommand{\subsubsection}{\@startsection {subsubsection}{3}{\z@}
{-2ex \@plus -0.1ex \@minus -.2ex}
{.2ex \@plus.2ex }
{\normalfont\small\sffamily\bfseries}}                        
\renewcommand\paragraph{\@startsection{paragraph}{4}{\z@}
{-2ex \@plus-.2ex \@minus .2ex}
{.1ex}
{\normalfont\small\sffamily\bfseries}}

\newcommand{\@mypartnumtocformat}[2]{%
\setlength\fboxsep{0pt}%
\noindent\colorbox{ocre!20}{\strut\parbox[c][.7cm]{\ecart}{\color{ocre!70}\Large\sffamily\bfseries\centering#1}}\hskip\esp\colorbox{ocre!40}{\strut\parbox[c][.7cm]{\linewidth-\ecart-\esp}{\Large\sffamily\centering#2}}}%
\newcommand{\@myparttocformat}[1]{%
\setlength\fboxsep{0pt}%
\noindent\colorbox{ocre!40}{\strut\parbox[c][.7cm]{\linewidth}{\Large\sffamily\centering#1}}}%
\newlength\esp
\newlength\ecart
\def\@part[#1]#2{%
\ifnum \c@secnumdepth >-2\relax%
\refstepcounter{part}%
\addcontentsline{toc}{part}{\texorpdfstring{\protect\@mypartnumtocformat{\thepart}{#1}}{\partname~\thepart\ ---\ #1}}
\else%
\addcontentsline{toc}{part}{\texorpdfstring{\protect\@myparttocformat{#1}}{#1}}%
\fi%
\startcontents%
\markboth{}{}%
{\thispagestyle{empty}%
\begin{tikzpicture}[remember picture,overlay]%
\node at (current page.north west){\begin{tikzpicture}[remember picture,overlay]%
\fill[ocre!20](0cm,0cm) rectangle (\paperwidth,-\paperheight);
\node[anchor=north] at (4cm,-3.25cm){\color{ocre!40}\fontsize{220}{100}\sffamily\bfseries\@Roman\c@part}; 
\node[anchor=south east] at (\paperwidth-1cm,-\paperheight+1cm){\parbox[t][][t]{8.5cm}{
\printcontents{l}{0}{\setcounter{tocdepth}{1}}%
}};
\node[anchor=north east] at (\paperwidth-1.5cm,-3.25cm){\parbox[t][][t]{15cm}{\strut\raggedleft\color{white}\fontsize{30}{30}\sffamily\bfseries#2}};
\end{tikzpicture}};
\end{tikzpicture}}%
\@endpart}
\def\@spart#1{%
\startcontents%
\phantomsection
{\thispagestyle{empty}%
\begin{tikzpicture}[remember picture,overlay]%
\node at (current page.north west){\begin{tikzpicture}[remember picture,overlay]%
\fill[ocre!20](0cm,0cm) rectangle (\paperwidth,-\paperheight);
\node[anchor=north east] at (\paperwidth-1.5cm,-3.25cm){\parbox[t][][t]{15cm}{\strut\raggedleft\color{white}\fontsize{30}{30}\sffamily\bfseries#1}};
\end{tikzpicture}};
\end{tikzpicture}}
\addcontentsline{toc}{part}{\texorpdfstring{%
\setlength\fboxsep{0pt}%
\noindent\protect\colorbox{ocre!40}{\strut\protect\parbox[c][.7cm]{\linewidth}{\Large\sffamily\protect\centering #1\quad\mbox{}}}}{#1}}%
\@endpart}
\def\@endpart{\vfil\newpage
\if@twoside
\if@openright
\null
\thispagestyle{empty}%
\newpage
\fi
\fi
\if@tempswa
\twocolumn
\fi}

\newif\ifusechapterimage
\usechapterimagetrue
\newcommand{\thechapterimage}{}%
\newcommand{\chapterimage}[1]{\ifusechapterimage\renewcommand{\thechapterimage}{#1}\fi}%
\def\@makechapterhead#1{%
{\parindent \z@ \raggedright \normalfont
\ifnum \c@secnumdepth >\m@ne
\if@mainmatter
\begin{tikzpicture}[remember picture,overlay]
\node at (current page.north west)
{\begin{tikzpicture}[remember picture,overlay]
\node[anchor=north west,inner sep=0pt] at (0,0) {\ifusechapterimage\includegraphics[width=\paperwidth]{\thechapterimage}\fi};
\draw[anchor=west] (\Gm@lmargin,-9cm) node [line width=2pt,rounded corners=15pt,draw=ocre,fill=white,fill opacity=0.5,inner sep=15pt]{\strut\makebox[22cm]{}};
\draw[anchor=west] (\Gm@lmargin+.3cm,-9cm) node {\huge\sffamily\bfseries\color{black}\thechapter. #1\strut};
\end{tikzpicture}};
\end{tikzpicture}
\else
\begin{tikzpicture}[remember picture,overlay]
\node at (current page.north west)
{\begin{tikzpicture}[remember picture,overlay]
\node[anchor=north west,inner sep=0pt] at (0,0) {\ifusechapterimage\includegraphics[width=\paperwidth]{\thechapterimage}\fi};
\draw[anchor=west] (\Gm@lmargin,-9cm) node [line width=2pt,rounded corners=15pt,draw=ocre,fill=white,fill opacity=0.5,inner sep=15pt]{\strut\makebox[22cm]{}};
\draw[anchor=west] (\Gm@lmargin+.3cm,-9cm) node {\huge\sffamily\bfseries\color{black}#1\strut};
\end{tikzpicture}};
\end{tikzpicture}
\fi\fi\par\vspace*{270\p@}}}

\def\@makeschapterhead#1{%
\begin{tikzpicture}[remember picture,overlay]
\node at (current page.north west)
{\begin{tikzpicture}[remember picture,overlay]
\node[anchor=north west,inner sep=0pt] at (0,0) {\ifusechapterimage\includegraphics[width=\paperwidth]{\thechapterimage}\fi};
\draw[anchor=west] (\Gm@lmargin,-9cm) node [line width=2pt,rounded corners=15pt,draw=ocre,fill=white,fill opacity=0.5,inner sep=15pt]{\strut\makebox[22cm]{}};
\draw[anchor=west] (\Gm@lmargin+.3cm,-9cm) node {\huge\sffamily\bfseries\color{black}#1\strut};
\end{tikzpicture}};
\end{tikzpicture}
\par\vspace*{270\p@}}
\makeatother

\usepackage{hyperref}
\hypersetup{hidelinks,colorlinks=false,breaklinks=true,urlcolor= ocre,bookmarksopen=false,pdftitle={Title},pdfauthor={Author}}
\usepackage{bookmark}
\bookmarksetup{
open,
numbered,
addtohook={%
\ifnum\bookmarkget{level}=0 
\bookmarksetup{bold}%
\fi
\ifnum\bookmarkget{level}=-1 
\bookmarksetup{color=ocre,bold}%
\fi
}
}

\newcommand{\AR}{$\mathcal{AR}\;$}

\newcommand{\MC}{$\mathcal{MC}\;$}
\newcommand{\CDR}{$\mathcal{CDR}\;$}

\begin{document}


\begingroup
\thispagestyle{empty}
\begin{tikzpicture}[remember picture,overlay]
\coordinate [below=12cm] (midpoint) at (current page.north);
\node at (current page.north west)
{\begin{tikzpicture}[remember picture,overlay]
\node[anchor=north west,inner sep=0pt] at (0,0) {\includegraphics[width=\paperwidth]{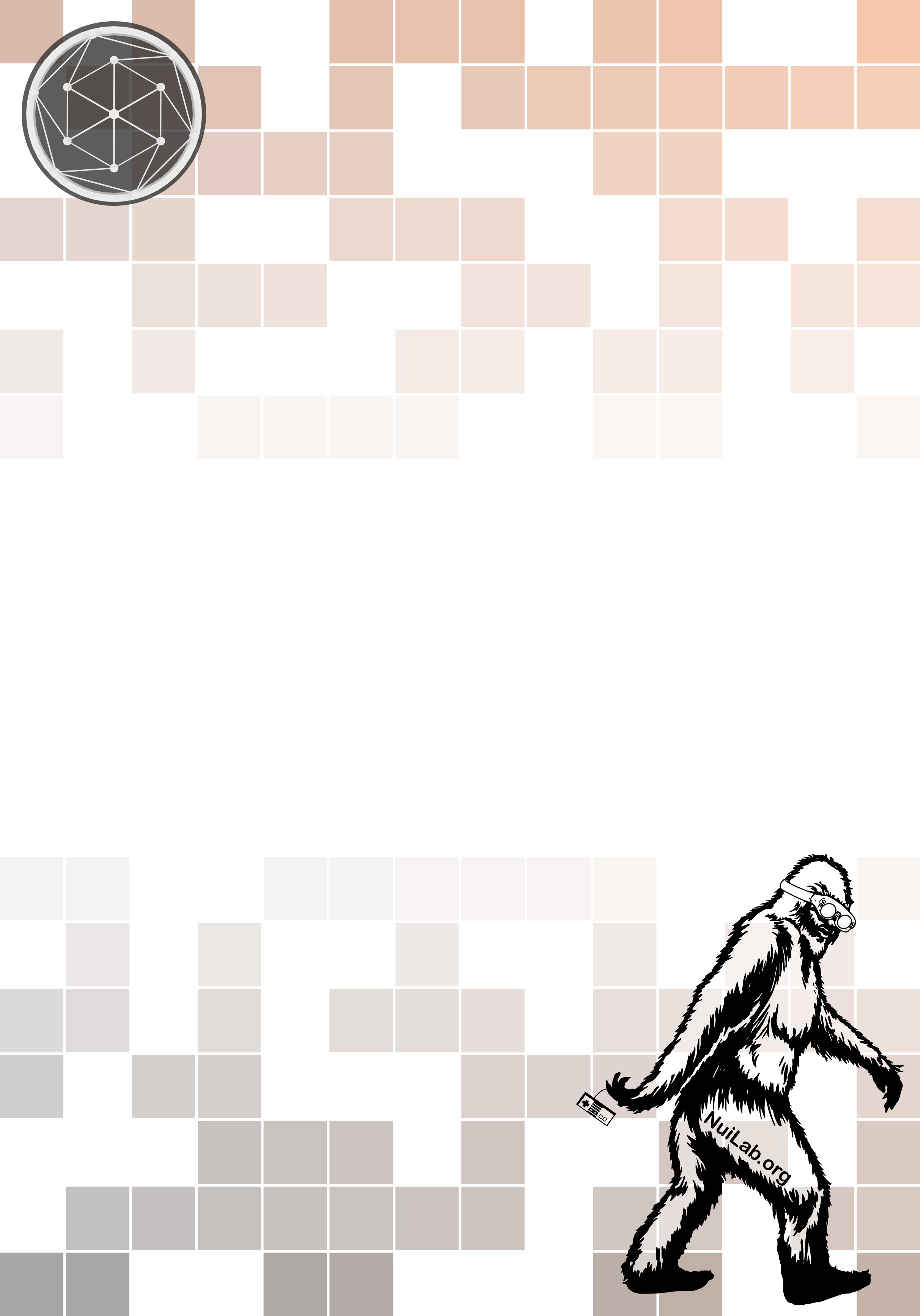}}; 
\draw[anchor=north] (midpoint) node [fill=ocre!30!white,fill opacity=0.6,text opacity=1,inner sep=1cm]{\Huge\centering\bfseries\sffamily\parbox[c][][t]{\paperwidth}
{\centering A Concise Guide to Elicitation Methodology\\[15pt] 
{\Large For Human-Computer Interaction Students and Beyond}\\[20pt] 
{\huge{\centering Adam S. Williams}}

{\huge{\centering Dr. Francisco R. Ortega}}}}; 
\end{tikzpicture}};
\end{tikzpicture}
\vfill
\endgroup

\newpage
~\vfill
\thispagestyle{empty}

\noindent Copyright \copyright\ 2021 Adam S. Williams \& Francisco R. Ortega\\ 

\noindent \textsc{Published by Adam Williams \& Francisco R. Ortega}\\ 



\noindent Licensed under the Creative Commons Attribution-Non Commercial-No Derivatives 4.0 International (the ``License''). You may not use this file except in compliance with the License. You may obtain a copy of the License at \url{https://creativecommons.org/licenses/by-nc-nd/4.0/legalcode}. Unless required by applicable law or agreed to in writing, items distributed under the License is distributed on an \textsc{``as is'' basis, without warranties or conditions of any kind}, either express or implied. See the License for the specific language governing permissions and limitations under the License.\\ 

\noindent \textit{First printing, May 2021} 


\newpage
\thispagestyle{empty}

\hspace{0pt}
\vfill
\noindent \textbf{Dedication:}\hfil \break  \hfil \break  

Adam S. Williams --- I would like to dedicate this guide to my father James who taught me to work both smarter and harder. 
\hfil \break  \hfil \break

Francisco R. Ortega --- I dedicate this work to Dr. Wim B\"{o}hm, Professor at Colorado State University, forever in my heart. 

\vfill
\hspace{0pt}

\newpage
\thispagestyle{empty}
\hspace{0pt}
\vfill
\vfill
\hspace{0pt}

\newpage
\thispagestyle{empty}

\hspace{0pt}
\vfill
\noindent \textbf{About the Authors:}\hfil \break  \hfil \break  

Adam S. Williams is a Ph.D. candidate in computer science at Colorado State University with a focus on human-computer interaction. Adam works in the Natural User Interaction lab under Dr. Francisco R. Ortega. His research is on multimodal interaction design for augmented reality head-mounted displays. This work has included elicitation studies and input comparisons geared towards finding the right inputs for natural feeling interactions stereoscopic virtual environments. Recently his work has been examining multimodal interactions in information-rich environments (Immersive Analytics). 

\hfil \break  \hfil \break

Francisco R. Ortega, Ph.D., is an Assistant Professor at Colorado State University (CSU) and Director of the Natural User Interaction (NUI) lab (https://nuilab.org). His research focuses on multimodal and unimodal interaction (gesture-centric), covering both gesture recognition and elicitation (e.g., a form of participatory design). Improving user interaction by (a) multimodal elicitation, (b) developing interactive techniques and (c) improving augmented reality visualization techniques lies at the heart of his research. Dr. Ortega is currently co-PI for the CSU DARPA's CwC project where leads the Faelyn Fox transition project to prototype a system to aid child-development assessment and PI and coPI from National Science Foundation, among others. In August he will start a new project for ONR developing a framework and model for AR visualization systems. Dr. Ortega has published over 60 peer-reviewed articles including in top venues of his field (IEEE TVCG journal, ACM ISS, IEEE VR, etc.), has published two books, and has been awarded 5 US patents. 

\vfill
\hspace{0pt}

\newpage
\thispagestyle{empty}
\hspace{0pt}
\vfill
\vfill
\hspace{0pt}

\newpage
\thispagestyle{empty}

\hspace{0pt}
\vfill
\noindent \textbf{Acknowledgments:}

First, and foremost, we would like to acknowledge the grant from the Colorado Department of Higher Education's OER Council that was a big part of this initiative and that keeps supporting faculty (\url{http://masterplan.highered.colorado.gov/oer-in-colorado/}). 

We would like to thank all of our sponsors, the National Science Foundation (NSF), and Defense Advanced Research Projects Agency (DARPA) for their continued support of this work.

This work was supported by the NSF National AI Institute for Student-AI Teaming (iSAT) under grant DRL 2019805. NSF also supported this work with the grants IIS 1948254, IIS 2037417, CNS 2016714, CNS 2106590, and BCS-1928502. This work was further supported by the DARPA contract number W911NF-15-1-0459.

The opinions expressed are those of the authors and do not represent views of the NSF, DARPA, ONR, or any other agency.

Adam S. Williams would like to thank his co-author Francisco Ortega for the opportunity to work on this book, for his guidance on research, and his continued facilitation of Adam's professional development and growth. He would like to also thank all of the reviewers and researchers that have provided feedback, both good and bad, to his work. This feedback has helped him to both improve as a researcher and as an author.

Francisco R. Ortega would like to thank his Ph.D. student Adam Williams, who's work with the natural user interaction (NUI) Lab provided much of the text in this guide and his amazing work. Francisco would also like to acknowledge all of his present and past students in the NUI Lab, his former Ph.D. advisors Armando Barreto and Naphtali D. Rishe, and his close collaborators in CSU including Bruce Draper, Ross Beveridge, Nathaniel Blanchard, Nikhil Krishnaswamy, Chris Wickens, Ben Clegg, and Cap Smith. Finally, He wants to give his gratitude to the 3DUI \& HCI community, including Doug Bowman, Rob Teather, Wolfgang Stuerzlinger, Anil Ufuk Batmaz, Mark Billinhurst, Victoria Interrante, Danielle Szafir, Scott Mackenzi, Amy Banic, and Lisa Anthony, among others.  

We would also like to thank all of the people that helped to provide images for this book. The NUILAB logo (cover page - Yeti) was provided by Adam S. Williams, the other NUILAB logo on the top left of the cover was provided by Alain Galvan, and the chapter images were proved by Ian Marshall and Adam S. Williams.  

\vfill
\hspace{0pt}

\newpage
\thispagestyle{empty}
\hspace{0pt}
\vfill
\vfill
\hspace{0pt}

\chapterimage{Teasser.png}
\chapter*{Preface}

Input systems such as gestures, speech, and other emerging interaction modalities are the next evolution of natural user interactions for stereoscopic head-mounted displays. Society has been at this turning point before. In 2007, Apple introduced the iPhone, changing the landscape of interaction by making multi-touch a pervasive input method. Apple, like others, relied on the research conducted by the many scientist before them (see ~\cite[Ch.~8]{ortega2016interaction} for a brief summary of multi-touch research). This change in interaction paradigm from the traditional windows icon menu points (WIMP) to multi-touch interactions allowed many people to become part of the digital revolution that may have been left out had WIMP been used instead. 
Today, while we see Oculus Quest 2 and other virtual reality (VR) head-mounted displays increasing their user base, they have not  yet become as pervasive as the multi-touch display. While multi-touch displays still have many open research questions left, VR, and in particular augmented reality (AR) optical-see through headsets (e.g., Microsoft HoloLens 2) still need much work. The appropriate interactions for these technologies are unknown and there is a major difference between interactions with HMDs and multi-touch devices. Interaction in a VR or AR headset are not necessarily intuitive in all cases. AR glasses are expected to become the interactive device of the future, utilized at the workplace, schools, and day to day life. If these glasses are developed with difficult to use interaction techniques their use may be avoided by segments of the population, thus leaving many persons behind. Interactions with these devices must be studied to aid this coming interaction and technological paradigm shift. 

This guide is born out of the need to have a correct knowledge of elicitation studies. This is of particular importance. While it may be tempting for some to write up an elicitation study to derive some additional publication out of existing observational data, elicitation studies, if done correctly, can provide deep insight into human interaction patterns and facilitate the development of natural user interaction systems. Additionally, as elicitation's use continues to grow, questions about its own methodologies are called into question. This guide provides the foundation needed to be able to continue the efforts to improve user interactions using elicitation research methodologies. We hope that this will serve as a open source of information for anyone that wants to incorporate it in their teaching and research. Finally, the authors hope that this will serve as the start of a 3D multimodal interaction book. 



\chapterimage{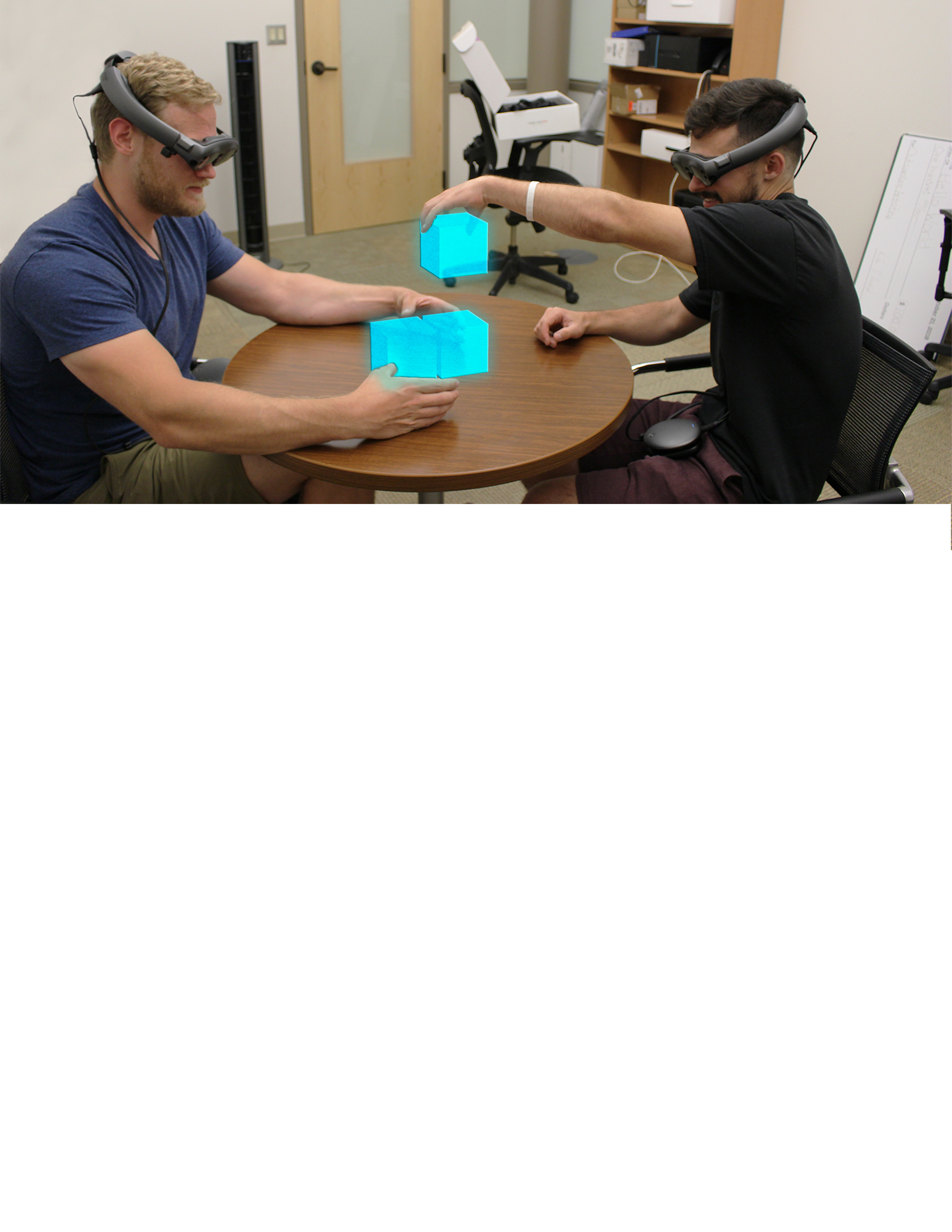} 

\pagestyle{empty} 

\tableofcontents 

\cleardoublepage 

\pagestyle{fancy} 




\chapterimage{Robot.png} 
\chapter{Introduction}

Consider the impact that the desktop computer, smartphone, and tablet have had on people's lives. Next consider the transitions of interaction techniques and interaction modalities used across those devices. The desktop computer used a keyboard along with a mouse, where a laptop emerged with a touch-pad in place of the mouse. Of course you can still use the mouse with the laptop, but as the form-factor of these devices changes so to do the ways that we interact with them change. Currently cellphones with multi-touch displays are common place. In multi-touch devices there is still a virtual keyboard but there are also options for swipe to text, speech to text, and many other novel interaction techniques. 

As novel technologies continue to emerge newer interaction techniques and modalities must be designed to match them. Virtual reality (VR), a system where a user dons a head-mounted-display (HMD) that replaces the real world with a virtual display came first with a 6-degrees-of-freedom (6DoF) controller then later with gesture interactions. Augmented reality (AR) is one of the key technologies expected to become pervasive in the not too distant future. AR is similar to VR except that the user can now see virtual content superimposed on the real world. AR devices commonly ship with gesture and speech interactions in place of that 6DoF controller.

One of the open questions in the field of interaction design is "what inputs or interaction techniques should be used with AR devices?" The transition from a touch-pad and a keyboard to a multi-touch device was relatively small. The transition from a multi-touch device to a HMD with no controllers or clear surface to interact with is more complicated. This book is a guide for how to figure out what interaction techniques and modalities people prefer when interacting with those devices. The name of the technique covered here is \textbf{Elicitation}. Elicitation is a form of participatory design, meaning design with direct end user involvement. By running an elicitation study researchers can observe unconstrained human interactions with emerging technologies to help guide input design. 

\section{Elicitation overview}

Knowing the user is a big step towards designing usable systems~\cite{HAN71}. A technique used to discover what users' interactions with an existing or emerging technology would be is to run an elicitation study~\cite{VIL+20}. Elicitation studies can help researchers to observe users interactions and thoughts. Often these studies are done using an emulated version of emerging technology~\cite{VAT+13, PIU+13}. Some other uses of elicitation have included conceptual~\cite{CHA+16} and existing~\cite{KHA+19} technologies.

Elicitation was popularized in human-computer interaction (HCI) research by Wobbrock et al. in 2005~\cite{WOB+05}. Latter the same team used elicitation methodology for multi-touch inputs~\cite{WOB+09}. That study was described as a ``guessability study'' because of it's goal of finding interactions that were highly discoverable to new users of a multi-touch system~\cite{WOB+09}. Part of this observational methodology's design stems from distributed cognition~\cite{HOL+00}. Elicitation works by removing the gulf of execution form a system so that a participant can interact with it in any manner they choose, ideally allowing their natural interactions with that system to occur. The gulf of execution includes any barriers to interactions with that system. Examples of these barriers include poor recognition systems, a bad user interface, and un-intuitive interaction techniques. When enough users are observed naturally interacting with a system their interactions can be aggregated to derive a set of user defined interaction techniques. 


Elicitation is a form of participatory design, that is, the participants of the study are helping to design the interaction techniques to be used with the examined technology. As would be expected from studies leveraging participatory design, one of elicitation's primary goals is often to develop a set of aggregated user interactions~\cite{COH+12, BUC+13}. That set of interactions is referred to as a consensus set. Elicitation can also be a rich source of observational user data. This observational data often leads to discoveries that are more generalizable than a consensus set alone would be. These discoveries have included the impact of scale on interaction generation~\cite{TAR+18, PHA+18}, the timing information around co-occurring gesture and speech inputs~\cite{ISS, LEE+08, TVCG}, user modality preference when multiple modality options are available~\cite{CAR+05, MOR12, NEB+14}, or that users prefer multimodal interactions more as task cognitive load increases~\cite{OVI+04}.


Elicitation study design is quickly becoming a popular research tool. This is demonstrated by the widely varied technology spaces that have used elicitation to help guide input design. Some of these domains include internet of things home set ups~\cite{ZAI+15}, computer-aided design~\cite{KHA+19},  mobile devices~\cite{RUI+11}, mid-air gestures~\cite{ORT+17, TVCG, PIU+13}, multi-touch surfaces~\cite{MIC+09, BUC+13, ORT+17}, and multimodal television browsing~\cite{MOR12, NEB+14}. Other novel uses of elicitation include using multi-touch devices to control physical objects through virtual representations of said entities~\cite{GUP+17} and imposing physical constraints on users to help establish interaction sets suitable for impaired and non-impaired persons~\cite{ALT+16, RUI+15}.

These studies represent only a small portion of the elicitation studies that have been run. As of  more than $216$ elicitation studies have been run with $5,458$ participants executing $3,625$ commands~\cite{VIL+20}. With this widespread use comes a stream of modifications and improvements to the original 2005 study design. One of the early changes occurred ten years after the original paper when the agreement index, a metric of participant agreement, was changed to be the agreement rate~\cite{WOB+05, VAT+15}. More recent changes include metrics for between group comparisons~\cite{VAT+15}, production agreement~\cite{VAT+15}, speech proposal consensus metrics~\cite{MOR12}, statistics to help identify the impact of chance agreement~\cite{TSA18}, and measures of the dissimilarity of proposals~\cite{VAT+19}.

Some studies still use the original methodology~\cite{WOB+09, VIL+20} while others have entirely altered the process~\cite{CAF+18}. There have been changes to the presentation of referents~\cite{VIL+20}, the Wizard of Oz systems used~\cite{COH+08}, and techniques used to prime users with a certain mindset~\cite{CHA+16, RUI+15} or mental frame~\cite{CAF+18}.

Most commonly elicitation has been used in unimodal input scenarios, where developers examine a single input modality. The most examined input modality was multi-touch then a few years later it became mid-air gesture. That said, elicitation has been used for a wide range of unimodal gesture based inputs from foot gestures to head gestures~\cite{VIL+20}. As new technologies are emerging, elicitation is starting to see use as a tool for multimodal input design. Examples of this are seen in studies examining gesture and speech based inputs~\cite{MOR12, KHA+19, TVCG, ISS, NEB+14}.

\chapterimage{Observation.png} 
\chapter{Observational Studies}

A very powerful tool for Human Computer Interaction (HCI) research is observational study. Observational studies are user studies where people are viewed while interacting with a system. A framework for viewing actors (people) in their environments is called ``Distributed Cognition'' (DCog)~\cite{HOL+00}.

Distributed cognition is a framework for the conceptualization of how actors interact with their environments proposed by Hollan et al. in 2000~\cite{HOL+00}. More specifically it is a framework for analyzing the way that cognition is shared across multiple actors or subsystems. Traditionally cognition has been viewed as a process that happens internally to one actor when interacting towards the desired outcome with a system. An example of this is seen in HCI research when an actor, in this case, a participant in a study, is presented with the action (referent) ``move left’’. The desired outcome is that the virtual object displayed should be translated to the left. The actor's cognitive process is the way that they are internally devising the execution of that action. In a more abstract sense, the unit of analysis when examining the cognitive processes involved in this task is an individual actor. 

In DCog the unit of cognition, meaning the unit of analysis, is outside of the individual user. Next, cognition happens both inside and outside of the brain. This raises the question; ``what is the cognitive process of one actor with this system and how is the process that the system has impacting that actor?'' DCog can be extended to ask ``what cognitive framework do multiple actors in a collaborative setting use with a system.'' The relationship between actors and their environment under DCog is viewed as bidirectional. The environment influences the way actors use the system and the actor impacts the was the system is used.

Cognitive processes involve the transmission and transformation of information. These processes have paths of information transmittal that when viewed in the aggregate form an architecture. This is called ``socially distributed cognition~\cite{HOL+00}’’. With this architecture, we can observe how individual cognitive processes transform when transitioning from an individual to group settings, and how group cognitive processes transform when transitioning to individual processes. We can also examine how participation in a group impacts individual actor's cognitive processes. At a high level, people establish and coordinate structure in their environment. This takes effort to maintain. People off-load cognitive effort whenever possible and there are improved load balancing dynamics in social organization~\cite{HOL+00}.

People are not passive actors in their environments. Between a person and their environment, many levels of interactions are at play. DCog labels this feature ``embodied cognition~\cite{HOL+00}.'' The processes in these interactions include an actor's memory, time, attention, and external resources. The important finding here is that the parts of an actor's environment are not passive stimuli. Actor's interactions with a system are bidirectional. An example is seen when considering the manipulation techniques implied by a coffee cup's shape. The way an actor picks up a coffee cup is different from the way they would pick up a plate. The object has an impact on the way that it is interacted with. 

\section{Affordances}

When an object has strong visual indicators of how it should be interacted with those indicators are referred to as perceived affordances~\cite{NOR88}. Affordances are an important component of embodied cognition. They can lower the amount of time it takes a user to learn how to interact with the object in question by making the interaction technique immediately present. Examples can be seen in light switches. A switch affords that it can be placed in an on or off position thus turning the light on or off. A knob implies that it has many radial positions to be placed in, implying a dimming feature. Affordances can be learned. Norman says that the affordance of a telephone is making calls which has remained consistent between models of phones but is not immediately obvious upon visual inspection~\cite{NOR88}. Affordances have been shown to impact how co-occurring gestures and speech are created~\cite{MAS+16}. Most often if something has a strong affordance, that affordance will be what is gestured about for that object~\cite{PIN+10}. This interaction means that actions are context-dependent on the activity being done. Actions are formed by both internal and external resources.

\begin{exercise}
Pick three objects in your room or office. What are the affordances of those objects? Example: my water-bottle has a loop that affords attachment and a snap lid that affords closing and opening. 
\end{exercise}

\section{Development of distributed cognition}

DCog framework is built upon previous observational studies done by Holland and Hutchins. Most commonly they mention a study done with airline pilots~\cite{HUT95}. In this study, they were interested in why a new air tape display was not well received by pilots. The new display was digital, an advancement from the previous electromechanical display. The digital display had abilities that went beyond the previous air tape. Even so, pilots expressed a preference for the older display. By observing the pilots interacting with the old system they found that the pilots not only used the speed information given (correctly added to the new display) but also the relative locations of stimuli on the screen (lost in the new display). This finding provided evidence that a system impacts how an actor uses it, and that observational studies are useful for uncovering nuanced interaction information. 

Systems use symbols to represent some nature of the system. A folder icon on a computer represents a container for files. The observation of the pilots utilizing symbols on the display as references without connection to the symbols meaning led to a new category of events being introduced by DCog.  Symbols are also tokens that have meaning inside of themselves. The position of the folder icon next to a recycling bin could symbolically serve as a reminder to trash the folder's contents. The new class of events is events in which a representation's structure but not its reference is changed. This lends to analyzing the ways that people offload cognitive effort by utilizing the structure of their environment. When people use tokens to aid in executing an interaction, it can be seen as the intelligent use of space. A multi-touch surface study found that users often drag shapes to the edge of the screen to return to the later~\cite{WOB+09}, exhibiting this intelligent use of space.

The pilots' preference for the older display is an instance of experts, people who were trained on a previous system, being used to test a new system. An issue to consider when doing this sort of study is the amount of bias that experts bring with them to the new system. In an experiment where both expert and novice users were asked to produce mid-air gestures for interactions with an anesthesia system~\cite{JUR+18}. The experts were anesthesia doctors and the novices were medical students early in their training. The gestures produced by experts were derived from the gestures they would use with the actual machine. The existing machine used a knob to adjust the flow of anesthesia, the experts gestured use of an invisible knob. The novices produced gestures that were less tied to the existing system.

\section{Legacy bias in observational studies}

When interaction techniques are influenced by previous systems they are said to have legacy bias~\cite{MOR+14}. An example of a legacy gesture is using a two-finger pinch as seen in touch screens with a mid-air gesture system. The gesture system could accept two hand expansions (i.e. two hands starting at a central point and expanding outwards) because of its extended recognition capabilities. Legacy bias is a double-edged sword. legacy gestures are typically more memorable because users have experience using them, or variations close to them, in the past~\cite{MOR+14}. This is a benefit in interaction design which could be exploited. The risk is that legacy gestures may not fully utilize the capabilities of the new system, or that they are not as ergonomically sound as the new system can allow for. There is more information on Legacy Bias located in Chapter~\ref{CH:FurtherReading} Section~\ref{SEC:LegacyBias}.

\section{Culture}

An important confounding variable to observe in DCog is culture. Culture was previously accepted as being formed by actors. DCog adds that actors are formed by culture. This means that An actor's culture can provide tools for interactions but can also add biases to the interactions. These bias’ can include blindness to potential solutions.  Most semaphoric gestures are culturally biased (e.g., thumbs up).  

\chapterimage{Legos.png} 
\chapter{Gestures}

Traditionally, elicitation studies in HCI have examined gesture interactions. The following sections give a basic understanding of gestures, their classification, and their formation.

\section{The mechanics of gesturing}

The foundations of gesturing lie in the mechanics which set them to motion. The framework commonly used in HCI is the ``kinematic chain model’’~\cite{BAL+00}. proposed by Guiard in 1988~\cite{GUI87}. The `` kinematic chain model’’ is a framework for viewing how bi-manual interactions are executed by humans. This model is built out of observational studies done by Guiard~\cite{GUI87}.

When building this model only right-handed people were observed. This allowed for high inter-subject homogeneity of hand preference. The cost was a loss of model generalizability to left-handed persons. There is an unstated assumption that the model generalizes but is the reverse for left-handed people. 

\section{Categories of manual activity}

The principles of this framework are stemmed from the three categories of manual activity. The first category is uni-manual interactions: actions done by one hand. By their nature uni-manual interactions are asymmetric. The second category is asymmetric bi-manual: movements that involve the usage of both hands. However, each hand's contribution to the task is different. The last category is symmetric bi-manual: movements where both hands work in unison. 

Guiard argues that there are no tasks that can be considered asymmetrically uni-manual. When examining tasks commonly viewed as asymmetrically uni-manual such as writing, one hand is writing, the other hand is framing the paper. Another uni-manual task is throwing a dart. Guiard argues that we cannot know if the opposite hand is providing some sort of assistance in balance or momentum. Even assuming that no assistance to the act of throwing is provided, the opposite hand is preforming the action of doing nothing. With these observations, the category of asymmetrically uni-manual is collapsed into the asymmetric bi-manual task category.  

This new framing of manual activity changes the way that handedness is viewed. Being right-handed implies that someone chooses to use their right hand for uni-manual tasks. As uni-manual is now bi-manual the term hand preference is used. People will exhibit a lateral preference for the types of tasks that they assign their hands. People will show manual superiority at tasks when using their preferred hand for that task. 

At the time this article was written (1988) there was a gap in the literature of neuropsychology. An argument was taking place about whether people had ``brainedness'' or not. That is to say, do people favor one hemisphere of their brain over the other or not. This attribute was correlated to being right or left-handed. Guiard's new framework allowed researchers to use it regardless of their stance of brainedness because it removed the handedness from the equation. People merely had hand preference.

\subsection{Principles of motion}

In manual action, there exists a ``right to left spatial reference in manual motion.’’ The right-hand finds its reference in the left hands positioning. Consider the task of sowing, the left-hand positions the fabric while the right does the stitching. This means that the right hand's accuracy is interwoven with the left hand’s actions. This point was elaborated by a study of handwriting done by Guiard~\cite{GUI87}. In this study, a carbon pad was placed under a sheet of paper. The task was for a participant to write an essay on the paper. The resulting data from this experiment was the essay written on the paper and the carbon papers pressure etches from the process of writing the essay. The results showed that the left hand re-frames and positions the paper while the right-hand inserts the content into the paper. The carbon paper had a much smaller window of writing than the full sheet of paper due to the re-positioning and the writing was done at an angle. This is taken to mean that the left hand will provide the framing for the object and the right hand will act upon that object.

There also exists a ``left to right contrast in the spatial-temporal scale of motion’’. Meaning the left hand will exhibit a relatively low temporal frequency of movements and high spatial movements. The right hand will exhibit more quick and short movements. This notion is reinforced with Weber’s Law~\cite{HOA+30}. Weber’s law states that a motor makes a compromise between slow large movements and short rapid movements. Mirrored here where each hand represents one end of that compromise. The left hand is specialized in providing large relatively slow actions whereas the right has been specialized to provide short quick content insertion activities. This work references an unpublished study run by Guiard where people were asked to turn knobs with their left and right hand. The effect of these knobs was a cursor moving on a television screen. Participants were prompted to move the cursor randomly with the left knob controlling the y-axis and the right knob controlling the x-axis. The results show large slow y-axis oscillations and small quick x-axis oscillations. 

The last observation is that there is a ``left-hand precedence in action.’’ This builds upon the previous two observations. The left hand by necessity acts first and the right hand follows. The left hand must position the paper before the right can insert content. 

\begin{exercise}
To personally observe the ``left to right contrast in the spatial-temporal scale of motion’’ write a half page of text on a sheet of paper. As you are writing pay attention to how you position the page. Most likely, when you start to move down more than a quarter of a page you will re-position the page with your non-dominate hand to make writing more easy for your preferred hand. For a more salient example of this try the same task with one hand behind your back. 
\end{exercise}

\subsection{The kinematic chain model}

The culmination of these observations is a model of manual activity called the ``kinematic chain model’’ which is used to view bi-manual cooperation of gestures. This model explicitly views the hands as simple motors. Motor being defined as a unit that creates motion. Thus, the hands (motors) are treated as black boxes with no regard for their internal workings. 

This kinematic chain model states that the hands can be viewed as motors placed in series. The output of the left motor (n) provides the input to the right motor (n+1). Motor n acts first. Motor n must have greater inertia than motor n+1. This feature is ignored when the hands act independently. These motors represent a hierarchical structure that facilitates different spatial and temporal frequencies of movement. This is not to say that the kinematic chain must have only two motors, only that it must have at least 2 motors and end with the right hand. With all of these, the opposite is the case with left handed persons, meaning that right hand will frame the tasks of the left hand. 

\subsection{Bi-manual gestures}

Bi-manual commands have been found to be more memorable and more enjoyable than uni-manual commands, particularly in the case of abstract commands~\cite{DEL+19}. In that study, The hands' labor was divided into activation and control gestures which mirror the spatio-temporal preferences shown in the kinematic chain model. Bi-manual interactions are better for both experienced and inexperienced users, with greater gains for inexperienced users~\cite{BUX+86}.  This study had 13/14 subjects immediately start by using both hands. Two-handed interactions outperformed one-handed interactions in each group (expert by 15\%, novice by 25\%). The experts using one hand outperformed novices using one hand by 85\% whereas experts using two hands vs novices using two hands only outperformed by 32\%. No statistical significance was found between experts using one hand and novices using two hands. 

A recent study using the Microsoft Hololens 1 tested user performance on bi-manual gesture interactions~\cite{CHA+18}. This study tasked 38 participants with preforming basic object manipulations in Augmented Reality (AR). The time it took to match a virtual object to a displayed template and the relative error of that placement were dependent variables. This study assumed that a hand was left or right based on the sign of the cross product of that hand and the participant's head position. This meant that if a user crossed their arms the left hand would be registered as the right hand. 

Participants would make an ``L'' gesture (thumb out, index straight) to signal they were ready to begin. They then used the input interface provided to perform the object manipulations. Participants were primed with indicators of the rotations.  A $4^{th}$ degree polynomial scaling ratio was used on continuous gestures. They tested five interaction techniques:``spindle + rise'', ``arc-ball'', ``wire-frame'', ``hands locking into gesture'', ``hands starting positions'', and ``Hands locking into gesture''.

Participants were told to complete as many interactions as possible in six minutes. They found the ``hands locking into gesture'' and ``wire-frame'' methods had the highest user enjoyment and the lowest user frustration. The accuracy of the interactions was similar across all techniques.  All techniques outside of the ``spindle + raise'' had high gesture interaction similarity. Some interaction techniques were duplicated or minimally modified. This could have caused highly similar interaction times between interaction methods. The main difference between ``hands starting positions'' and ``hands locking into gesture'' was the time that a gesture is registered. In ``hands locking into gesture,'' a user must pinch to start an interaction, in ``hands starting positions'' a user only had to align their hands in the desired area of the screen. 

This study shows that bi-manual interactions can be beneficial in AR environments~\cite{CHA+18}. The participants in this study chose hands locking into gesture as preferred 42\% of the time (next highest 26\%). The results should be viewed carefully due to the highly similar nature of the interactions tested. 


\section{Gestures as communication}

Prior to being researched as an input device in the field of HCI, gestures were extensively examined within the context of human to human communication. The following sections detail that research at a high level. Knowing how gestures are used in day to day life can help developers to best design intuitive interaction techniques that utilize common gesture themes and dynamics from everyday life. 

\subsection{A taxonomy of gestures}

The first step towards understanding human gestures is to understand the different taxonomies that they can fit into. Most taxonomies in use today evolved out of the one proposed by Efron in 1941~\cite{EFR41}. This taxonomy contains five categories, listed below:

\vspace{\baselineskip}

\begin{vocabulary}[Emblems]
Gestures that convey meaning such as a thumbs up. Emblems are typically specific to a culture.
\end{vocabulary}

\begin{vocabulary}[Icons]
Gestures whose movements provide a metaphor of an object or action. Tracing a table or visually displaying a ball bouncing with one’s hand are examples of icons.
\end{vocabulary}

\begin{vocabulary}[Metaphors]
Metaphors are like icons, but the thing being gestured about is abstract.
\end{vocabulary}

\begin{vocabulary}[Deictics]
Deictic localize a position in space. Pointing at a place and saying ``there''. Deictics are normally difficult to interpret unless accompanied by speech.
\end{vocabulary}

\begin{vocabulary}[Beats]
Beats are simple rapid and repetitive movements. Typically, not synchronous with speech. They can emphasize parts of the speech.
\end{vocabulary}

\vspace{\baselineskip}

These categories are expanded by McNeil in something he calls``Kendon's Continum''~\cite{MCN92}. This gesture taxonomy brakes gestures down into slightly more granular categories.

\vspace{\baselineskip}

\begin{vocabulary}[Gesticulation]
Motion that embodies meaning and can be tied to accompanying speech.
\end{vocabulary}

\begin{vocabulary}[Speech-framed gestures]
A gesture that has a place in a sentence and completes that sentence.
\end{vocabulary}

\begin{vocabulary}[Pantomime]
A gesture or sequence of gestures that presents a narrative.
\end{vocabulary}

\begin{vocabulary}[Sign]
A gesture with a commonly accepted lexical meaning. American sign language gestures are signs.
\end{vocabulary}

\vspace{\baselineskip}

\textbf{Emblems} were unchanged from Efron~\cite{EFR41}. Under this taxonomy ``gesticulations'' and ``speech framed gestures'' are further expanded into the following categories.

\vspace{\baselineskip}

\begin{vocabulary}[Iconic]
Gestures that present a concrete image of an entity or action.
\end{vocabulary}

\begin{vocabulary}[Metaphoric]
Gestures that depict abstract concepts.
\end{vocabulary}

\vspace{\baselineskip}

\textbf{Deictic} and \textbf{Beats} were used unchanged from Efron~\cite{EFR41}. 

\vspace{\baselineskip}

It is important to note that McNeil's taxonomy comes from the field of linguistics which is very interested in how co-speech gestures appear in human communication. Not all of these types of gestures are used when interacting with a system. Quek elaborated on McNeil's work with an emphasis on creating a gesture interaction language for use with computers~\cite{QUE94}. Building on Quek's and McNeil's work a taxonomy of gestures adapted for HCI was introduced by Karam and Schrefel in 2005~\cite{KAR+05}. This taxonomy is shown below.

\vspace{\baselineskip}

\begin{vocabulary}[Manipulative]
Gestures that control some entity by applying a nearly 1 to 1 relationship with that entity.
\end{vocabulary}

\begin{vocabulary}[Semaphoric]
Any gesture that employs a stylized dictionary of hand positions or movements to convey a message.
\end{vocabulary}

\begin{vocabulary}[Gesticulation]
Co-verbal gestures.
\end{vocabulary}

\begin{vocabulary}[Language gestures]
Gestures correspond to a lexical meaning.
\end{vocabulary}

\begin{vocabulary}[Multiple gesture styles]
A chaining of the other types of gestures.
\end{vocabulary}

\vspace{\baselineskip}

\textbf{Deictic} was used unchanged from Efron~\cite{EFR41}.

\vspace{\baselineskip}

The most recent adaptation of a gestural taxonomy comes from Aigner et al~\cite{AIG+12}. This is the State-of-the-art in gesture classification.

\vspace{\baselineskip}

\begin{vocabulary}[Pointing]
Formerly deictic gestures.
\end{vocabulary}

\begin{vocabulary}[Stroke]
Concentrated flick movements.
\end{vocabulary}

\begin{vocabulary}[Pantomimic]
Gestures describing an action.
\end{vocabulary}

\begin{vocabulary}[Iconic]
Gestures describing a shape.
\end{vocabulary}

\begin{vocabulary}[Manipulation]
Gestures designed to guide movements with a one to one correspondence.
\end{vocabulary}

\vspace{\baselineskip}

\textbf{Semaphoric} was used unchanged from~\cite{KAR+05}.

\vspace{\baselineskip}

Both Aigner et al's taxonomy and Karam and Schrefel's taxonomy include sub-categories based on the movements of a gesture~\cite{AIG+12, KAR+05}. These sub-categories are ``static'' which are stationary gestures, ``dynamic'' which are moving gestures, and ``strokes'' which are concentrated flick movements.

To establish their taxonomy Aigner et al. ran a user study with 12 participants. Participants were recruited from varied industries to reduce bias. Participants were grouped in pairs. One participant was shown a referent (action to be completed) and the other was given an object. The first participant's task was to instruct the other how to execute the referent shown by gesturing. The participants were in different rooms and could see each other with video feeds. They could not hear each other.  The referents were intentionally low-level actions not often present in human to human communication. 

Using the human on the other screen may have influenced participants to use more human to human communication gestures. To reduce that bias an alternative approach would use a virtual avatar on the other end of the screen. The study only used participants from one culture~\cite{AIG+12}. That might have impacted the gestures produced by making them culturally specific (semaphoric gestures would be affected the most). The goal of the study was to categorize types of gesture proposals at a high level and not to create a gesture interaction set, thus, cultural bias may not have had a major impact on their proposed taxonomy. Semaphores vary by culture but are still binned in as semaphores. 

\subsection{Phases of gesturing}

Individual gestures can be broken down into five phrases. There is an initial phase where the body is at rest. Followed by a preparation phase where motion starts. Then a stroke phase, which contains most of the gestures information content. The stroke is typically the peak in physical effort of the gesture. Holds are the static phases that may precede or follow the stroke. Lastly, there is a recovery phase where the body is returned to the rest position~\cite{MCN05}. These phases are important when considering co-occurring speech gestures. Each phase lines up with a different structural element of an utterance. A stroke normally lines up with the subject of a sentence~\cite{KEN80, MCN05}. These phases have been used to segment gestures which can aid in creating gesture recognizer systems~\cite{MAD+16, ISS}. 

\subsection{The importance of gestures and speech}\label{sec:why_GS}

Interface design should be intuitive~\cite{NIE+04}. Outside of HCI knowledge, there is a vast collection of human to human communication literature. Human to human interactions are arguably what we as people are most familiar with. Multimodal systems that can mirror human to human interactions can reduce the amount of learning needed for technologies' use. Gestures have been shown to be deeply linked with speech. Gestures and speech together constitute language~\cite{MCN05}. Combined they create an integrated system for speech production and comprehension~\cite{KEL+10}. This has been shown in many ways. Kelly et al. had participants identify images that most closely represented short videos. The videos were of someone performing an action such as chopping chives on a cutting board. The images contained a word and a gesture that was either strongly or weakly related to the image. When the images were less congruent with the video performance was degraded~\cite{KEL+10}. That study was the foundation of the ``Integrated-systems hypothesis,'' which states that gestures and speech have a bidirectional influence on each other and have obligatory interaction~\cite{KEL+10}. This work reinforced Kendon and McNeil's conclusions that gestures and speech are synchronous modes of cognition and co-expressive~\cite{KEN80, MCN05}. People even gesture when no one is looking, such as when speaking on telephones~\cite{RIM82}.

Goldwin Medow et al. expanded on the importance of gestures in communicative tasks by conducting an experiment where participants were asked to describe the steps needed to complete a math problem. The conditions were whether or not the participants were allowed to gesture~\cite{GOL+01}. They found that when gesturing participants more quickly and more accurately worked through the steps of the math problems. This implies that participants' cognitive load was lessened by allowing the offloading of mental processes to other memory domains, such as from the short-term verbal memory to the visual-spatial sketchpad. 

Co-occurring gestures and speech have several desirable qualities. They can express meaning. They often contain information that is non-redundant to the accompanying speech~\cite{KRU+02}. Gestures typically have less error than speech which means that listeners can correct speech errors with them~\cite{MCN92}. Gesturing helps facilitate lexical access~\cite{KRA98}. When gesturing for ``the ball rolled’’ the gesture for \textit{roll} is considered to activate the lexical retrieval of the word roll. Making the word easier to access, which as seen in Goldwin Medow et al., can make tasks easier~\cite{GOL+01}. Complementary to that, when lexical access is more difficult gesturing frequency increases~\cite{MOR+04}.  Additionally, when gestures are prohibited speech production is slowed~\cite{RAU+96}. 

Lastly, gestures facilitate language development. When synchronized with speech, gestures aid infants in interpreting the meaning of unknown words~\cite{ZAM+11}. Infants will start to gesture as communication before they begin to speak. These communication gestures increase in complexity as speech narratives complexity increases~\cite{COL+10}. This supports the notion that co-speech gestures develop in parallel with linguistic ability. 

This work emphasizes the notion that gestures and speech are a deeply bound and rich mode of communication. Some people have gone so far as to say gesturing constitutes a visual language~\cite{ARD+14}. The work of linguistics on co-occurring gestures and speech has been and continues to be leveraged by HCI researchers. Bourguet et al. showed that the timing of deictic gestures is often within -200ms to 400ms of the beginning of an expression~\cite{BOU+98}. This lines up with work done in this lab (Natural User Interaction Lab), which found that gestures occur with a median value of 130ms before the co-occurring speech~\cite{ISS}. Timing windows based on Bourguet et al.'s results have been leveraged to improve recognizer accuracy~\cite{EIS+04}. Research has tapped into the alignment of gestures with structural sentence elements found by Kendon~\cite{KEN80} to improve interaction systems~\cite{SHA+08}. Other work in HCI has confirmed that the keyword relevant to the desired manipulation occurs close in time to the gesture produced, which can be leveraged to further hone a recognition window for multimodal systems~\cite{OVI+97}. The likelihood of gesturing is greatly increased based on the type of interaction requested~\cite{FEY+99}. Abstract interactions are more likely to be done with speech whereas placement is often done with gestures. By taking the lessons learned in linguistics and viewing them through an HCI lens we can develop better models of interaction and better recognition systems. 

\chapterimage{ReferentSideBySide.png} 
\chapter{Elicitation Background}

The following chapter outlines terms used in elicitation studies. Later it use the work done by Wobbrock et al. in $2009$ as an elicitation example~\cite{WOB+09}.

\textbf{Refresher}: elicitation is a type of participatory study design that can help us map inputs to actions for emerging technologies. Outside of input design alone, elicitation is a tool for helping researchers understand user behavior. One such insight is that some users will prefer larger motions for large objects than small objects when completing the same command~\cite{TAR+18, PLA+17}. Another behavioral contribution from an elicitation study is that users have a preference for upper-body gestures when a whole-body system is available~\cite{ORT+19}. Elicitation studies are typically carried out using some variant of the original study's methodology~\cite{WOB+05, WOB+09} along with the metrics from the later papers~\cite{VAT+15, VAT+16}.

\section{Terms}

\begin{vocabulary}[Elicitation]
Prompting potential end-users of a system to generate inputs for that system.
\end{vocabulary}

\begin{vocabulary}[Agreement]
A measure of how many participants proposed the same interaction for a given command (referent).
\end{vocabulary}

\begin{vocabulary}[Proposal (sign)]
The input suggested by a user for a given command (referent).
\end{vocabulary}

\begin{vocabulary}[Referent]
The command/action which the input proposal will execute.
\end{vocabulary}

\begin{vocabulary}[Consensus set]
A set of highly agreed upon interactions.
\end{vocabulary}

\begin{vocabulary}[Wizard-of-Oz (WoZ)]
A study design where a system's recognition capabilities are emulated by an experimenter.
\end{vocabulary}

\begin{vocabulary}[Think-aloud]
When participants are asked to describe what lead to the formation of their input proposal.
\end{vocabulary}

\begin{vocabulary}[Binning]
Partitioning proposals into equivalence classes based on pre-defined metrics (i.e. the number of fingers used, hand posture, motion).
\end{vocabulary}

\section{Elicitation protocol}

Commonly elicitation studies will use $25$ (Median = $20$ Standard Deviation (SD) = $4$) participants~\cite{VIL+20}. These participants will be presented with referents and asked to generate a input proposal for each. Sometimes participants will be requested to generate unique proposals for each referent~\cite{PIU+13}, other times they may be permitted to propose the same interaction across multiple referents~\cite{TVCG, ISS}. The chosen referents will normally be clustered around an intended use domain such as ``generic object manipulations in AR''~\cite{TVCG}. The referents will be shown to the participants one at a time in random order. Participants may be encouraged to use think aloud protocol when generating proposals. Importantly, elicitation studies often use a Wizard-of-Oz (WoZ) experiment design to remove the gulf of execution between the participant and the system by having the researcher act as the systems recognizer allowing participants to believe they are interacting with a live system~\cite{WOB+09}.

Elicitation studies often use video to capture participant interactions~\cite{VIL+20, VOG+18}; however, sometimes other means such as skeletal tracking~\cite{VAT+19, NEB+14} and touch data ~\cite{WOB+09} are used. The participants interaction data is then paired with the observational data from interviews and any participant commentary from the experiment or the think-aloud data. Video data is hand-annotated by one or more raters with descriptions for the gestures produced by participants recorded in detail~\cite{WOB+09, ORT+19, TVCG, ISS}. At this level the gestures will be very granular with notes on things like fingers used, hand positions, the proposal's semantic features~\cite{SUK+18}, and the motion of the gesture~\cite{ISS, PIU+13}. These granular bins are then merged into equivalence classes bases on predefined features such as gesture similarity or the number of fingers used.  For example, some studies have shown that people do not differentiate between 1-2 finger mid-air gestures and 3-4 finger mid-air gestures~\cite{CHA+16}. With that in mind it is appropriate to bin one finger left to right swipe with two-finger left to right swipe.e~\cite{WOB+09, CHA+16}. When skeletal data is collected computer vision techniques can be used to bin gestures eliminating the need for hand annotation and potentially the human bias introduced during hand annotation~\cite{VAT+19}. The binning of proposals is an important step towards removing the individual-level characteristics of the proposals in favor of a more generalizable consensus set.

Agreement metrics are used to measure consensus of the binned interaction proposals by referent. Often this agreement is measured using the \textit{Agreement Rate} formula~\cite{WOB+09}. Agreement rate and other commonly used consensus metrics are covered in Chapter~\ref{CH:Metrics}. Based on the outcome of these metrics, a set of consensus gestures can be proposed for referents that achieve above a predetermined score in the used metric. For agreement rate this level is often around $0.3$ for sample sizes of $20$~\cite{VAT+15}.

\subsubsection{Example case}

Wobbrock et al. asked participants to interact with objects on a large touch-screen table. The chosen domain was surface computing and as such referents included \textit{move a little}, \textit{move a lot}, \textit{pan}~\cite{WOB+09}. There were 27 referents in total. Twenty paid participants were asked to complete the study with a mean age of 41.3 (sd = 15.6). These participants created 1080 proposals for interactions. Wobbrock et al. used a think-aloud protocol and collected video data.~\cite{WOB+09}. The resulting proposals were binned based on their movement~\cite{WOB+09}. The results of Wobbrock et al.'s study was a consensus set of user-defined multi-touch gestures and a taxonomy of gesture use~\cite{WOB+09}.

\section{Gesture and speech elicitation}

There has been limited work on co-occurring gesture and speech elicitation. In most gesture+speech elicitation studies researchers gave participants a list of acceptable speech and gesture commands then observed the participants using them in a virtual environment. Hauptmann and Hauptmann et al. used dietic gestures to rotate, translate, and scale objects on a 2d screen~\cite{HAU89, HAU+93}. Mignot et al. used 8 participants and found that speech was used more abstractly than gestures~\cite{MIG+93}. Robbe presented some preliminary work showing that constrained speech dictionaries can be quickly learned without being frustrating to end users~\cite{ROB98}. Anastasiou et al. is a more recent WoZ study on gestures and speech, however, their gestures were touch-based gestures. This list is exclusively studies that use WoZ methodology to find co-occurring gesture and speech input patterns and usage

Morris in 2012~\cite{MOR12} elicited speech and gesture inputs for mid-air interactions with a large wall-mounted television. This was a WoZ study with 22 participants (11 groups). The participants were asked to sit on a couch in pairs and to interact with a web browser on the television. A Microsoft Kinect was mounted on top of the television and participants were told that it could recognize their gestures and speech perfectly. Participants were prompted with referents such as ``refresh page'', or ``new tab.'' This study represents one of the most well-rounded speech elicitation studies done.

With this design, they found that participants typically found that one modality was best for certain tasks, such as speech for choosing a movie. These modalities were individual choices and not always the same between people. Highlighting the importance of having multiple available inputs. Often people would propose speech and gesture synonyms for tasks. 

Another crucial advancement in elicitation methodology made by Morris's work was the introduction of two new metrics specific to speech elicitation~\cite{MOR12}. These being ``consensus distinct ratio'' and ``max consensus''. These are aimed at capturing the agreement rates and goodness of fit for elicited speech proposals. $\mathcal{AR}$(r) falls apart when used for speech because it has far more potential proposals than gestures. An utterance could be ``place that there'' or ``place there'' or ``that there.'' What would be the appropriate bins when selecting equivalence classes for these commands? The waters get very murky when speech is involved. The max-consensus metric is the percent of participants suggesting the most popular proposed interaction for a referent. The consensus distinct ratio is the percent of distinct proposals given for a referent. These new metrics allow us to gauge the most guessable proposals as well as gauge the potential complexity of a referent. Guessable gestures having a higher max consensus and difficult referents having a higher consensus-distinct-ratio.

\section{Elicitation criticisms}

Elicitation studies have received criticism in a few areas. First is the impact of legacy bias on input proposals~\cite{MOR+14}. Legacy bias is when a gesture proposal is heavily informed by participants' interactions with prior technology. An example is a participant saying ``F5'' when suggesting a speech input for refreshing a browser page~\cite{MOR12}. This bias could be leveraged and is not always considered a negative quality of input proposals~\cite{ORT+17, KOP+15}. Using legacy or near legacy interactions can lead to a more discoverable interaction set as it can mirror users' preconceived mental models of interactions~\cite{KOP+15}. Several methods exist for reducing legacy bias, further widening the variances found in elicitation procedures. 








Another elicitation concern is the issue of chance agreement which occurs when the input proposal space is small enough that high agreement rates could conceivably be caused by random chance because the agreement rate formula assumes an infinite space of gesture proposals where that space may be more limited~\cite{TSA18}. Tsandilas (2018) suggests that participants will  cluster around a subset of gestures making the actual proposal space sampled from much more limited~\cite{TSA18}. A way to resolve this is to calculate the Fleiss' Kappa coefficient and the associated chance agreement term to assess the impact of chance agreement~\cite{TSA18}. 

A debated question is whether a system should be designed by an expert or by users (user-centered design).  One clear advantage of user-driven gestures is the evidence found by Nacenta et al. that elicited gestures are up to 24\% more memorable~\cite{NAC+13}. It should be noted that these were user-defined and tested at the individual level (user \textit{a} made a set of gestures then was tested on that set). More relevantly Morris et al. found that users preferred user-defined gesture sets at the aggregate level~\cite{MOR+10}. This was, however, a small effect. Wobbrock et al. found that user-defined gesture sets are more guessable (able to be discovered by users) than expert-defined sets~\cite{WOB+05}.

\section{Referent display}

The goal when presenting a referent is to establish the command to be completed by the input proposal. If eliciting commands for television-based web browsing then a referent would be \textit{refresh page}~\cite{MOR12}. \textit{Refresh page} could be presented as text reading ``refresh page'', an animation of a web page being refreshed, or an experimenter reading the referent aloud. In the case of Morris, $2012$, and Nebeling et al., it was both showing the effect of the referent (the animation) and stating its name aloud~\cite{MOR12, NEB+14}. Note that both of these studies used gesture and speech as input modalities. The effect of speech imitation can be seen in their results; however, it was never mentioned that reading the referents out-loud contributed to the overlap between spoken referents and participant speech proposals. 

Referents have been presented to participants in a variety of ways which becomes problematic when the elicited proposals may be highly impacted by the choice of referent presentation. Referent display techniques have included animations paired with spoken aloud instructions~\cite{MOR12}, images~\cite{SIL+15, CHE+18, ROV+14, LEE+14}, animations alone~\cite{ISS, DIM+16, HOF+16, KHA+19, MAY+16, KOU+16}, text alone~\cite{TVCG}, only read aloud~\cite{CON+13}, text and animation~\cite{FEL+18, ORT+19}, text and read aloud~\cite{ORT+17, RUI+15, ZAI+15}, and the combination of text, reading aloud, and animations~\cite{VAT+13}. On occasion, the exact form of display is left slightly unclear~\cite{CHA+16}. With this wide range of prompts used and some evidence suggesting that referents can bias proposals~\cite{VIL+20}, we believe that further study of the impacts/implications of referent display are merited.

\section{Forms of elicitation}

There exist several common forms of elicitation~\cite{VIL+20}. The highest level categories of elicitation are ``open'' and ``closed''. In an open elicitation study participants are invited to generate any proposals they deem appropriate for the referent presented. In a closed elicitation study participants are often given a set of appropriate gestures to sample from when making proposals. The end goal of these two designs is somewhat different. When any command it permitted the goal is to find what participants would naturally choose to do. When a closed design is used the goal is to determine which gestures or interactions out of the ones given are best suited for the referents.

\chapterimage{JasonScreens2.png} 
\chapter{Designing An Elicitation Study}
\label{CH:Design}

This chapter outlines the design choices to be made when developing an elicitation study.

\section{Technology used}

One of the first choices made when designing elicitation studies is the choice of the technology and domain the elicitation will cover. This is one of the most important choices made during the design of an elicitation study. You should ask yourself ``has it been done before'', ``why should it be done'', ``will it's findings contribute to an under explored area'', and other questions around the need and novelty of your intended study. Ideally, an elicitation study should be aimed at finding something that is as of yet undiscovered. This finding can be small, such as examining how multi-touch gestures produced by participants have changes in the years since Wobbrock et al.'s 2009 study. The finding can also be large such as ``what gestures do people use in AR video games''. 

Around the same time that the selection of what technology to use is made, the domain of interest should be established. Will this elicitation study examine a generic goal such as general object manipulations or will it be more targeted (i.e. photo manipulations). The pairing of the technology used and the domain used will impact the results, the write up, the generalizability of the study, and the usability of the results.

\begin{exercise}
\label{EX:ElicitTech}
Make a list of technologies and domain pairs that you are interested in. After the list has a few items go through and write what a novel elicitation goal for each one. Lastly, write down what potential hurdles would keep it from being novel. At the end of this activity try to have one technology and domain that are a promising place to start a novel elicitation study design with. 
\end{exercise}

\section{Input modalities examined}

The section of what inputs to examine is the next foundational choice to make. Are you going to look at mid-air gestures or mid-air pen use? Maybe multi-touch or speech is more appropriate to the chosen domain and technology. In this step consider the likely inputs that users will have for technology (e.g. multi-touch on cellphones, gesture on AR-HMDs). Also consider the form factors and ergonomics of the inputs selected and the domain used. Eliciting mid-air gestures for long term tasks will cause fatigue, where micro-gestures (e.g., gestures formed with fingers and palms) might be more feasible. If eliciting for bicycle gestures for a trip computer, gloves work well where large trackers do not.  

It is also possible to compare several input modalities. This has been approached in two ways. The input modalities can be examined individually and combined~\cite{ISS} or all at once~\cite{MOR12}. In the case of gesture and speech elicitation this would look like having a gesture, speech, and gesture+speech condition or a single gesture and/or speech condition. This choice will be impacted by the goal of the study. If the study wishes to see how people interact in each modality, or how those interactions change between modalities then individual comparisons are best. If instead, the study is examining which input modality users will gravitate towards for each referent, the single input condition should be chosen. 

\begin{exercise}
Make a list of input technologies that could be used with the technology/domain pair that you came up with in Exercise~\ref{EX:ElicitTech}. Try to have five or more options. For each one consider how well they match the likely use cases for the chosen domain. Will users be seated or walking? Will they want to carry a controller or will they be at a desk where they can have many controllers. Next, rank each item in terms of how well suited it is for your chosen domain. 
\end{exercise}

\section{Referents}

The referents chosen will be dependent on the domain used. Some common referents include rotation about each axis, translation on each axis, scaling, and selection~\cite{ISS, TVCG, PIU+13}. These might be limited to uniform scaling~\cite{TVCG} or allow for non-uniform scaling axis~\cite{PIU+13}. Some more domain specific referents might include cutting a photo, pasting an object, deleting an object, or changing the color of an object. Examples of domain based referents can be found for television browsing in the work of Morris 2012~\cite{MOR12} or computer aided design in the work of Khan et al., 2019~\cite{KHA+19}.

When selecting referents try to keep in mind that the experiment should take an hour or less to complete. If too many referents are chosen the experiment might be extended, thus risking losing participant engagement. Another consideration is the type of input modalities chosen. If there will be several independent input conditions using the referents the experiment will last longer than it would with a single input condition. 

Another critical choice in referent design is how to display the referents. Referent display techniques have included animations paired with spoken aloud instructions~\cite{MOR12}, images~\cite{SIL+15, CHE+18, ROV+14, LEE+14}, animations alone~\cite{ISS, DIM+16, HOF+16, KHA+19, MAY+16, KOU+16}, text alone~\cite{TVCG}, only read aloud~\cite{CON+13}, text and animation~\cite{FEL+18, ORT+19}, text and read aloud~\cite{ORT+17, RUI+15, ZAI+15}, and the combination of text, reading aloud, and animations~\cite{VAT+13}. The referent display can impact the results of the study. If text or spoken referents are used in speech elicitation most often participants will repeat the referent~\cite{ISS, TVCG, MOR12, NEB+14}. On the other side of the spectrum, animations can bias the movement of elicited gesture proposals. Be careful to select a referent presentation that minimally biases interaction proposals. 

\begin{exercise}
\label{EX:ElicitRefs}
Using the technology and domain pair you determined for Exercise~\ref{EX:ElicitTech} create a list of referents that should be elicited. Make sure that these referents cover the most common tasks done in the domain you chose. These referents can be listed in categories. These can be based on the type of action (i.e., translation, scaling) or their correspondence to the real world (i.e., abstract, physics based). 
\end{exercise}

\begin{exercise}
List out 3 ways that you could display the referents you decided on in Exercise~\ref{EX:ElicitRefs}. For each type of referent display list one pro and on con of using it (e.g., assuming Speech elicitation referents: displayed as text - PROP: most clear communication of referent goals, CON: it may bias the produced utterances).
\end{exercise}

\section{Data collection}

Without collecting the data from the experiment the results would be meaningless. Video data is often a great starting point for elicitation studies data collection. Video will capture both the audio and visual activity of the participant during the experiment. It is best to use several video capture devices simultaneously to ensure that there is protection against technical errors or out of frame movements. Skeletal data can also be beneficial as it can allow the use of machine learning techniques to bin the elicited proposals~\cite{VAT+19}. 

Other data sources include system logs which can help record the time each trial starts and participant responses to prompts. These prompts can be entry and exit surveys, established surveys (i.e., NASA Task Load Index), or interviews with the participants. These will be covered in more detail in Chapter~\ref{CH:Metrics}. The choices made here will impact the write up of the results. If interviews are used researchers may be able to include a participant feedback or thoughts section. If surveys are used things like user overall perceived workload for each input can be established. 

\begin{exercise}
How would you collect data for the elicitation study being developed in these exercises? Write a list of technologies and data-streams you can use for data collection. Next to each on write the type of data you hope to get (i.e., .mp4, .csv) and the goal for collecting that data (i.e., video: capture participant interactions). 
\end{exercise}

\begin{exercise}
Generate at least five questions that you would ask a participant before an elicitation study for the technology/domain you listed in Exercise~\ref{EX:ElicitTech}. Next, create a list of five or more questions that you would ask after the study. Some examples of questions are ``which hand do you preffer'', ``how often do you play with a game controller''.
\end{exercise}

\begin{exercise}
Think about any surveys you would use for your study. List them out and write why you think you should use them. What additional insight dose the inclusion of this data-point give?
\end{exercise}

\section{Pilot studies}

Before embarking data collection a pilot study should be run. A pilot study is a small study that uses the same design as the main study. Pilot studies can help ensure data quality, confirm that the instructions are understandable, and help to find any lingering bugs in the software being used. Doing this will allow the full experiment to run smoothly. Pilot studies can also be used to test differing design options to see how the results may be impacted. As an example you could test text referents against animated referents for gesture+speech elicitation~\cite{ISS}. It is not necessary to fully interpret the data from pilot studies. Often the data can simply be checked for quality and that it matches the expected data.

\begin{exercise}
\label{EX:Pilot}
Find a friend and ask them to participate in the elicitation study you have been designing in these exercises. This can be a high-fidelity activity where you develop a full elicitation environment or a low-fidelity environment where you use PowerPoint or another easy means of emulating an elicitation environment. With either approach, run the elicitation study and record the results. Upon completion ask for their feedback on what made sense and what didn't. Ask what they would change. Ask yourself what you think you could change. Did the pilot reveal anything that was unexpected about your study? \textbf{Note:} If you intend to publish this as part of a paper be sure to have a approved IRB protocol (See Chapter`\ref{CH:Design} Section~\ref{Sec:Ethics}).
\end{exercise}

\section{Modifications}

Modifications can change the aims of an elicitation study greatly. An example is using physical constraints to cause participants to generate gesture proposals that are more realistic to physically impaired users likely inputs~\cite{RUI+15}. Other modifications can be pairing participants~\cite{MOR+14, MOR12}, priming users~\cite{CAF+18}, or using production where users generate N+ proposals per referent~\cite{FLORIDALEGACY}.

To remove some of the biasing found in speech elicitation we recommend using a goals based elicitation methodology. This can be done by presenting users with high level goals then breaking their subsequent actions into action/interaction pairs. This could look like asking a participant to remove any county with less than 5 major cities from a scatter-plot of city data. From that the participant may perform interactions for selection, deletion, and quarrying data points. With this approach there is more post processing involved on the researchers side. The benefit is that instead of priming a user with animations or text (i.e., ``select that point'') the participant is acting towards a higher goal. Similar methods of observing goal completion are common in information visualization work~\cite{AMA+05,BON+06}. Another approach is to use time delays between the presentation of the referent and the proposal generation by the participant. This could utilize the quick decay rate that working memory has, causing participants level of priming caused by the referent to be lessened~\cite{MOR+06}.

\begin{exercise}
Do you think that your elicitation study could use some modifications? Write down any changes that seem like they could improve the outcomes that you are looking for. Afterwards, you can run the pilot study from Exercise~\ref{EX:Pilot} with the modifications that seem most promising. Did any interesting differences in results arise out of the modifications?
\end{exercise}

\section{Ethics}
\label{Sec:Ethics}

Elicitation studies are inherently human subjects research. With that make sure that a research protocol is established with your institutions internal review board or ethics committee. This will require preparation of a consent form that tells potential participants what the risks, benefits, goals, and high-level tasks are for the experiment. video data will require a release form stating that video is able to be collected. There may also be forms for safety procedures (i.e., COVID-19 safety~\cite{UXCOVID}). The research protocol will need to be established and approved before any data collection is underway.

\chapterimage{Metrics.png} 
\chapter{Elicitation Metrics and Analysis}
\label{CH:Metrics}

\section{Agreement index}

A separate but crucial analysis found in elicitation studies is a computation of the agreement among proposals. Agreement can be seen as the number of pairs in agreement over the total number of pairs. Wobbrock et al. proposed the ``Agreement Index'' $\mathcal{A}$(r) formula in their 2009 paper (Equation~\ref{eq:a_r})~\cite{WOB+09}.

\begin{equation}
  \centering
  \mathcal{A}(r)=\sum_{P_i\subseteq P} \left( \frac{|P_i|}{|P|} \right)^{2}
  \label{eq:a_r}
\end{equation}

For a fixed referent $r$, the Agreement Index $\mathcal{A}$(r) is calculated using equation~\ref{eq:a_r}, where $P$ is the set of all proposals for referent $r$, and $P_i$ are the subsets of equivalent proposals from P. To demonstrate how this formula works suppose that 20 participants propose gestures for a single referent $r$.

\[|P|=|\{A,B,C,...,T\}|=20\]

These gestures (proposals) are sorted into 3 equivalence classes:

\[P_i=\{\{A,B,C,...,O\},\{P,Q,R\},\{S,T\}\}\]

This corresponds to $|P_1|= 15$, $|P_2|= 3$, and $|P_3|= 2$.

 \[ \mathcal{A}(r)= \sum_{P_i\subseteq P} \left( \frac{|P_i|}{|P|} \right)^{2}= \left( \frac{15}{20} \right)^{2}+\left( \frac{3}{20} \right)^{2}+\left( \frac{2}{20} \right)^{2}= .595\]

\section{Agreement rate}

In the $\mathcal{A}$(r) (Equation ~\ref{eq:a_r}) formula if there are no proposals pairs (perfect disagreement) the result is still greater than zero. To fix this issue of a referent agreeing with it self the $\mathcal{A}$(r) (Equation ~\ref{eq:a_r}) was changed to the $\mathcal{AR}$(r) formula, now called the ``Agreement Rate''~\cite{VAT+15}. The Agreement Rate formula is shown in equation~\ref{eq:ar}. 

\begin{equation}
  \centering
  \mathcal{AR}(r)= \frac{ \sum\limits_{P_i\subseteq P}{|P_i| \choose 2} }{ {|P| \choose 2} }
  \label{eq:ar}
\end{equation}


To demonstrate the new formulation consider the same scenario from the $\mathcal{A}$(r) (20 proposals, 3 equivalence classes).

\[\mathcal{AR}(r)= \frac{ \sum\limits_{P_i\subseteq P}{|P_i| \choose 2} }{ {|P| \choose 2} } = \frac{15\cdot14}{20\cdot19} + \frac{3\cdot2}{20\cdot19} + \frac{2\cdot1}{20\cdot19}=.574\]

The result is now .574, the previous result was .595. The Agreement Rate will always be lower than the Agreement Index. The transition from the Agreement Index to the Agreement Rate is a linear transformation that preserves the ranking of results. Meaning that while the Agreement Rate will be lower the conclusions of studies using the Agreement Index will still be valid. Most importantly, these formulas provide rankings within a singular study. \AR is not made to be compared between studies. It is dependent on the count of participants used and proposal space found making any comparisons done between studies inaccurate unless the studies were done in a near identical manner. Larger numbers in one study do not imply better results than lower numbers in another study. 

Based on simulations of probability distributions for $\mathcal{AR}$(r) (Equation~\ref{eq:ar}) with various participant counts Vatavu et al. suggest that an  $\mathcal{AR}$(r) of 0.30 can be considered high agreement~\cite{VAT+15}.

\section{Dissimilarity-consensus}

The most recent metric proposed for calculating gesture consensus is the ``dissimilarity-consensus'' method~\cite{VAT+19}. This is a method proposed by Vatavu that is used to compute objective measures of consensus between users' gesture proposals. Similar to the  $\mathcal{AR}$(r) metric. The main difference between the methods is the binning of equivalence classes. In  $\mathcal{AR}$(r) this is done by human raters. Humans rating proposals adds subjectivity to the results. Vatavu argues that this subjectivity can vary the results of a study when different raters are used. Vatavu references binning large cat mimicking gestures as the same whether the participant stands or sits down. In practice, the bins are typically more refined (1 finger with 2 fingers in a swipe gesture)~\cite{CHA+16}. Even so, Vatavu's argument that there is some added subjectivity by using human raters is accurate. 

\begin{equation}
\label{eq:Dis}
C_{R}(\tau)=\frac{\sum_{i=1}^{N} \sum_{j=i+1}^{N}\left[\Delta\left(g_{i}, g_{j}\right) \leq \tau\right]}{\frac{1}{2} N(N-1)}[\cdot 100 \%]
\end{equation}

To remove this subjectivity Vatavu proposes that gestures should be compared between every participant pair based on their similarity of articulation. The formula used to achieve this is seen in equation~\ref{eq:Dis}. in equation~\ref{eq:Dis} $g_{i}$ : represents the gesture elicited from the $i^{th}$ participant to some referent ${R}$. $\Delta$ represents the dissimilarity function that returns a positive number, an example of a $\Delta$ function is dynamic time warping (DTW). $N$ is the number of participants. $C_{R}(\tau)$ can be a number between 0 or 100 where 0 is very conservative (harder to achieve consensus)  and 100 is very permissive (easier to achieve consensus). Gestures $g_{i}$ and $g_{j}$ are said to be similar if $\Delta$  $(g_{i},g_{j})$ $\leq$ $\tau$. 


\begin{equation}
C_{R}^{\star}(\tau)=\frac{\sum_{i=1}^{N} \sum_{j=i+1}^{N}\left[\zeta\left(\Delta\left(g_{i, t}, g_{j, u}\right) \forall t, u\right) \leq \tau\right]}{\frac{1}{2} N(N-1)}[\cdot \cdot 100 \%]
\label{eq:PDis}
\end{equation}

Vatavu extended this equation to work for production studies. In production studies, a participant is asked to propose more than one gesture per referent. The goal of this procedure is to limit legacy bias~\cite{MOR+14}. The modified formula is shown in equation~\ref{eq:PDis}. in this formula $g_{i. t}$ is the  $t^{th}$ proposal collected from the  $i^{th}$ participant for referent ${R}$. $t$ and $u$ index the gestures proposed by participants $i$ and $j$ for referent $R$. A new function $\zeta$ is added to take all the dissimilarity values for all the $t$ by $u$ combinations of $ g_{i, t}$ and $ g_{j, u}$ and return one value. $\zeta$ can be a min, max, or avg function. 

The constant $\tau$ can be set by the experimenter with a high $\tau$ allowing for high consensus rates and a low $\tau$ leading to low consensus rates. Setting $\tau$ is compared to the binning of equivalence classes when using the $\mathcal{AR}$(r) metric~\cite{WOB+09}. This paper suggests modeling $\tau$ after a logistic growth curve and viewing the results across all the tested values of $\tau$ is the most appropriate approach~\cite{VAT+19}. The goal should be to get a fitted growth curve with a $\alpha$ of $0.05$. 

Each referent for any level of $\tau$ will have a binary matrix containing the results of each participant pair ([$\Delta (g_i,g_j) \leq \tau $]). Multiple samples of $\tau$ can be combined into one matrix. A hill-climbing algorithm or a correlation clustering technique can be applied to the matrices to find the largest separable cluster of gesture proposals. This cluster will be the consensus set. 

To test this new methodology Vatavu conducted a full-body gesture elicitation study with children ages 4 to 6. There were 30 participants with half being male. A Microsoft Kinect v1.8 sensor was used to capture skeleton data. The skeleton data was prepared by re-sampling it to 25 frames per second (fps), normalizing the height of the participants, and translating each skeleton to the origin of the coordinate system. Using the results from this study Vatavu highlighted the plausibility of the ``dissimilarity-consensus'' method. It is important to note that these were large full-body gestures. They did not capture finger data. The ``dissimilarity-consensus'' method should extendable to hand gesture elicitation studies. 

\section{Chance agreement}

To calculate the level of chance agreement ($P_e$) within the elicited proposals the chance agreement term (Equation~\ref{eq:PE}) is used. This term stems from the calculation of Fleiss' Kappa~\cite{TSA18}. In Equation~\ref{eq:PE} $m$ is the total number of proposals, $n_{ik}$ is the number of participants proposing proposal $i$ in bin $k$, $n_i$ is the total number of proposals for proposal $i$. The term $\pi_k$ reflects the chance that a rater classifies an item into category $k$ based on the times that category has been used across the data. $q$ is the space of possible proposals. The \AR can be inflated by chance agreement if the total number of distinct gestures proposed during the study is low. The use of $P_e$ allows us to compare the \AR value with the level of chance agreement to determine if the \AR is inflated because of high levels of chance agreement.

\begin{equation}
  \centering
  p_{e}=\sum_{k=1}^{q\_} \pi_{k}^{2}, \quad \pi_{k}=\frac{1}{m} \sum_{i=1}^{m} \frac{n_{i k}}{n_{i}}
  \label{eq:PE}
\end{equation}

\section{Speech metrics}

Speech was analyzed using two metrics of agreement. The first is max-consensus ($\mathcal{MC}$). \MC is the percent of participants proposing the most common utterance proposal~\cite{MOR12}. If $12$ participants proposed the utterance ``move left'' for the referent \textit{move left} and $5$ propose ``left'', $2$ propose ``move'', and $1$ participant proposes ``sideways''  the \MC equals $60\%$. The second speech metric is the consensus-distinct ratio ($\mathcal{CDR}$). \CDR is the percent of proposals for a referent that has over a baseline of $1$ participants proposing them~\cite{MOR12}. In the above-mentioned proposal scenario, the \CDR is $75\%$. \MC and \CDR were averaged across referents to gauge the general level of difference in metrics between the two studies. 

These metrics capture the peak and spread of the speech proposal space~\cite{MOR12}. If a referent has a proposal with a high \MC, that proposal is considered discoverable to novice users of this system. Alternatively, a high \CDR means that a referent has a high amount of disagreement on the best choice of proposals for that referent between participants. These are not exclusive metrics. It is possible to have a referent with a single highly proposed interaction (i.e., a high \MC) and a number of proposals that are suggested by single participants (i.e., high \CDR). This would imply that there is a clear most common utterance but not a clear second place or alternative choice utterance. By comparing these metrics and the top choice utterances for the speech proposals from the speech-alone and the gesture+speech input conditions, the differences in the elicited speech proposals across the two choices of referent display can be assessed.

\section{Consensus sets}

Commonly elicitation studies will result in the creation of a consensus set of common interactions. This set can be generated in several ways. A common choice is to take the top proposal for each referent and present it as the best interaction for that referent. The level of these interaction's fit with their associated referents will depend on the \AR achieved by that referent. If a referent has a low \AR (i.e., less than $0.1$) the most proposed interaction for that referent may not be very intuitive. This method of consensus set generation has been used across different interaction paradigms, including, multi-touch surfaces~\cite{ORT+19} and gesture interactions in AR~\cite{TVCG}. 

Sometimes it is beneficial to show more than a single most common interaction. Doing this can allow practitioners to build more robust systems, understand more of the entire interaction space observed, and to alias commands together to capture more of a novice users first choice interactions. An example of aliasing would be to make a swipe with two fingers and a swipe with the whole hand both work interchangeably for a given command. One approach to show a larger portion of the captured interaction space is to show the common movements along with the common hand poses used. This might look like saying that for the command ``move left'' the most common gesture was to use the right hand to push the object left with either a open palm or an extended index finger. This style of reporting has been done for some studies examining AR interactions~\cite{PIU+13, ISS}. This has also been seen in some multimodal studies where participants top few interactions across all modalities of input were shown~\cite{NEB+14, MOR12}. Another approach for presentation of a consensus set is to present the entire interaction set or all interactions proposed by more than some pre-determined baseline count of participants. In the past this has been achieved using heat-maps of the gesture space with a base line of two or more proposals needed~\cite{ISS}.

\section{Observations}

The types of observations gained during an elicitation study depend on the study design. Interviews, session videos, and think-aloud protocols all lead to increased observational data. Examples of this data include peoples preference for input modalities or gestures~\cite{MOR12}, noticing that people gesture a certain way when eliciting speech~\cite{TVCG}, and the way gestures are impacted by the scale of objects in the virtual environment~\cite{ORT+17, PHA+18}. These findings will typically use qualitative analysis and be mentioned as observations. 

To set up collection of interview data, determine before hand what questions you want to ask participants and add those to an interview list. After the experiment those questions will be asked first, with some time given to questions that arise out of watching the experiment. Doing this will establish constancy within the core questions. When watching the videos of the sessions be sure to note any interesting behaviors that occur to help with data interpretation. In some cases observations can be measured as a percentage of participants displaying that trait. As an example, five of the twelve participants mentioned that they preferred gestures over speech for simple object manipulations. 

\section{Surveys}

There are several surveys that can be used for elicitation studies. The most common ones are entry and exit questionnaires that are established by the researchers doing the experiment. These should include questions on demographics (i.e., age, gender, handedness, major/job) and on device experience (i.e., hours of weekly VR use). Knowing this will help the interpretation of participants interaction data. It might be that experienced VR users generate different inputs than people who have never used VR. Post-experiment surveys might ask participants to rank their interactions, rate the input modalities, or describe parts of their experience. Often participant experience or preference questions will use a Likert scale where options are given on a scale with a high and low end. An example Likert scale and question is ``I enjoyed mid-air gesture interactions'' with the responses 0: strongly disagree, 2: disagree, 3: neutral, 4: agree, 5: strongly agree. 

A commonly used survey is the NASA Task Load Index (TLX) which measures a participants overall perceived workload~\cite{HAR+88}. This is done by using a 21 segment scale question on 6 categories of workload then pairwise comparison questions for each grouping of those categories. The categories covered are mental demand, physical demand, temporal demand, performance, effort, and frustration. Upon completion a score for each category as well as an overall score is given. This survey can help to tell which inputs participant's found most difficult or least difficult to use. The sub category's scores can be beneficial when evaluating which interaction modalities are best suited or perceived to be the least frustrating in the examined environment. 

Surveys and questionnaires are normally analyzed using measures of central tendency (i.e., mean, median, mode, standard deviation) and often compared between each other using T-Tests or another statistical comparison appropriate to the type of data and normality of that data. When many comparisons are made be sure to adjust your alpha. In most cases a Bonferonni adjustment will work. Bonferonni adjustments are done by dividing your alpha value by the number of tests performed. This adjust the sensitivity of the alpha to accommodate for the the potentially large number of tests performed and chance significance.

\chapterimage{Elderly.png} 
\chapter{Further Reading}
\label{CH:FurtherReading}


\section{Ergonomics and input design}

In 2004 Nielsen et al. proposed a system for the development of ergonomic and intuitive gestures in HCI~\cite{NIE+04}. This method is often used in combination with the elicitation and WoZ methods proposed by Wobbrock and Vatavu~\cite{WOB+05, WOB+09, VAT+15, VAT+19}. Nielson et al. stated that usability has five main principles.

\begin{vocabulary}[Learnability]
Difficulty as measured by the time and effort it takes to reach a desired level of performance.
\end{vocabulary}

\begin{vocabulary}[Efficiency]
The mean performance of expert users.
\end{vocabulary}

\begin{vocabulary}[Memorability]
The ease of interaction use for novice users.
\end{vocabulary}

\begin{vocabulary}[Errors]
The rate of misused gestures.
\end{vocabulary}

\begin{vocabulary}[Coverage]
Operators (i.e. gestures) discovered vs total operators.
\end{vocabulary}

They also outline three human-based usability heuristics that developed interaction techniques should display:

\begin{itemize}
    \item Intuitive, easy to perform and remember
    \item Metaphorically and iconically logical towards functionality (i.e., does the interaction technique match the users mental model of how the interaction should work?)
    \item Ergonomic: not physically stressing for prolonged use
\end{itemize}

These can all be applied to any multimodal interface design. Recommendations for gesture interfaces include avoiding long static positions, repetitions, muscle tension, and forces on joints~\cite{NIE+04}. The first step towards finding a gesture set that fits under these criteria is to define the referents used. Then collect the proposals from users. After all the data is collected the input set should be extracted. This step includes binning into equivalence classes and finding agreement rates~\cite{WOB+09, VAT+15}. The last step is what sets this work apart from the previously mentioned elicitation methodologies. The collected input set should be put through three tests. First, users should be presented with the input(gesture) and asked to guess the command it is used for. Then, the memorability of the input should be tested by asking the user to perform the input when presented with a referent.  Lastly, the user should be asked to perform each input \textit{x} times and then report the overall level of stress for use. This process can be used for any input set creation. When the last three tests are used on an elicited input set that set can be determined to be ergonomic for the proposed interface. 

\section{Legacy bias}
\label{SEC:LegacyBias}

Legacy bias is a form of bias introduced into interaction proposals by participant familiarity with prior interaction techniques and technologies~\cite{MOR+14}. 

\begin{exercise}
If you were asked to zoom in using a mid-air gesture what gesture would you use?  
\end{exercise}

Most likely the gesture you proposed was the two-finger zooming gesture that is commonplace on touchscreen cellphones today. This gesture is an example of a legacy gesture, or a gesture that is transferred from a prior technology. Legacy gestures are not inherently bad, they can posses many beneficial traits. Some of those including memorability, knowledge transfer, and being easy for people to guess. The downside of legacy bias occurs when interaction proposals are not a good fit for the new environment that the elicitation is being done for. Perhaps mid-air gestures open up opportunities for gesture use that were not present with cellphones. If most gestures that are used are legacy gestures they may never realize the benefits offered by mid-air interactions. In between novel and legacy gestures are evolutionary gestures~\cite{UXEVO}. Evolutionary gestures are gestures that are heavily informed by past interactions but utilize a slightly different form that better matches the new environment in which they are used.

An elicitation study run on medical students and anesthesiologists does a great job of capturing this down side risk~\cite{JUR+18}. The study elicited mid-air gestures for the operation of anesthesia machines. This domain is very promising because any amount of physical contact in a operating room can increase the risk of spreading an infection. The downside is that anesthesia must be administered very precisely. If too much is given a patient can die, where too little could cause undue pain. 
During the study it was found that experienced anesthesiologists interaction proposals were highly biased by their legacy interactions with the current machines used for administering anesthesia. The gestures that the experts proposed included turning knobs and flipping switches as seen on the current machines. Alternatively the medical students who had no used those machines produced far more varied gesture proposals. That study did not assess if one set of gestures was better than the other. That said it is reasonable to believe that the novice produced gestures would be more discoverable to other novices than the expert produced gestures, especially if we consider that these novices may be trained without the currently used machines. In that case the legacy interactions are less memorable and less suited to this newer system of controlling anesthesia. That ill fit is the downside of legacy bias. 

\subsection{Legacy bias reduction}

Three methods for reducing legacy bias have been proposed~\cite{MOR+14}. These are partnered elicitation, priming, and production~\cite{MOR+14}. These methods have each seen use in elicitation studies~\cite{WIT+16, HOF+16, RUI+15, NEB+14}; however, the actual impact of their use has been seldom measured~\cite{FLORIDALEGACY}. 

\begin{vocabulary}[Priming]
Directly or indirectly influencing a participants mindset before running a study.
\end{vocabulary}

\begin{vocabulary}[Production]
Asking participants to produce N+ proposals per referent.
\end{vocabulary}

\begin{vocabulary}[Partnered Elicitation]
Grouping participants in pairs during the elicitation study. 
\end{vocabulary}

An example of priming is to ask participants to do jumping jacks before eliciting a proposal. The thought is that the activity may cause the participant to generate different and possibly more unique proposals. In the case of jumping jacks those proposals may be more full body than were the participant stationary before the elicitation study. Other ways that priming has been accomplished include weighted constraints~\cite{RUI+15} and having participants do kinesthetic activities~\cite{WIT+16, HOF+16}. Priming can also be more subtle such as a description of the types of gestures being elicited~\cite{CHA+16}.

Production is based on the idea that participants will exhaust their legacy proposals if asked for more than one proposal per referent. This would cause later proposals to be less biased. This has been found to be minimally effective for referents that are likely to invoke legacy biased proposals but not very effective otherwise~\cite{FLORIDALEGACY}. Production normally asks participants to generate three or more proposals per referent~\cite{CHA+16, LEE+14, KOU+16}.

When using paired elicitation participants are placed into small groups of two or three people. The goal is to cause increased variety in the elicited proposals by using collaboration. A few studies have been done using partnered elicitation; however, legacy bias reduction was not stated as the study's goal for using it~\cite{MOR12, NEB+14}. 


\chapterimage{Vydia.png}
\addcontentsline{toc}{chapter}{\textcolor{ocre}{Bibliography}}
\printbibliography


\cleardoublepage
\phantomsection
\setlength{\columnsep}{0.75cm}
\addcontentsline{toc}{chapter}{\textcolor{ocre}{Index}}
\printindex


\end{document}